# Ligand Directed Self-Assembly of Bulk Organic-Semiconductor/Quantum-Dot Blend Films Enables Near Quantitative Harvesting of Triplet Excitons

Jesse Allardice[a*], Victor Gray[a,b*], Simon Dowland[c*], Daniel T. W. Toolan[d*], Michael P. Weir[e*], James Xiao[a], Zhilong Zhang[a], Jurjen F. Winkel[c], Anthony J. Petty II[f], John Anthony[f], Richard Friend[a], Anthony J. Ryan[d], Richard A. L. Jones[e], Neil C. Greenham[a] and Akshay Rao[a†]

**Contacts**

[a]Cavendish Laboratory, University of Cambridge, J. J. Thomson Avenue, Cambridge, CB3 0HE, UK

[b]Department of Chemistry, Ångström Laboratory, Uppsala University, Box 532, SE-751 20 Uppsala, Sweden

[c] Cambridge Photon Technology, J. J. Thomson Avenue, Cambridge, CB3 0HE, UK

[d]Department of Chemistry, The University of Sheffield, Sheffield S3 7HF, United Kingdom

[e]Department of Physics and Astronomy, The University of Sheffield, Sheffield S3 7RH, United Kingdom

[f]Center for Applied Energy Research, University of Kentucky, Research Park Dr., Lexington KY 40511, USA

*: Authors contributing equally to this work.

†: Corresponding Author.

**Singlet fission (SF), an exciton multiplication process occurring in organic semiconductors,[1–6] offers a mechanism to break the Shockley-Queisser limit in single-bandgap photovoltaics.[7,8] If the triplet excitons generated by SF can be transferred to inorganic quantum dots (QDs), where they radiatively recombine, SF based photon multiplication is achieved, converting a single high-energy photon into two low-energy photons. Such a SF photon multiplication film (SF-PMF) could raise the efficiency of the best Si photovoltaics from 26.7% to 32.5%.[9] But a precise nanoscale morphology is required within such a film consisting of the appropriate morphology for the organic phase, allowing for efficient SF, within which the QD emitters are well dispersed on a tens of nm length scale to enable efficient harvesting of the triplets. However, it has been a long-standing problem that the individual components in organic-QD blends have a strong tendency to aggregate and phase separate during processing, due to a mismatch of their size, shape and surface energies.[10–12] Here, we demonstrate a QD surface engineering approach using an electronically active, highly soluble semiconductor ligand that is matched to the SF material, which allows us to direct the self-assembly process yielding solution processed films with well-dispersed QDs and minimal aggregation, as characterised by X-ray and neutron scattering and electron microscopy. Steady state and time-resolved optical spectroscopy show that the films support efficient SF (190% yield) in the organic phase and quantitative triplet energy transfer across the organic-QD interface, resulting in 95% of the triplet excitons generated in the organic phase being harvested by the QDs. Our results establish the SF-PMF as a highly promising architecture to harness the SF process to enhance PV efficiencies, and also provide a highly versatile approach to overcome challenges in the blending of organic semiconductors with QDs**

After decades of research and development, Si photovoltaics (Si-PV) are approaching the theoretical limit for power conversion efficiencies (currently 26.7% out of a possible 29.4% at 1 sun) as determined by the Shockley-Queisser limit.[7,13,14] One major factor determining this limit is the loss of energy from the conversion of high-energy photons to electrons, which produces significant amounts of wasted energy as heat. In 1979 D. L. Dexter proposed that the process of singlet exciton fission (SF), where a singlet exciton in an organic semiconductor decays into two independent triplet excitons, could be utilised to generate multiple electrons from a single high energy photon.[15] Thus, energy previously lost through thermalisation would become bound in a second excited state, circumventing this loss pathway and revising the absolute Shockley-Queisser limit upwards from 33% to 44% - a substantial gain.[8,9,16,17]

There are two principal approaches to incorporate a SF material with a Si-PV. The first of these is the generation of triplets from SF followed by their direct injection into the semiconductor via an electrical connection (as initially proposed by Dexter).[15,18,19] Such an approach, while promising, requires a change in cell design and thus has the added challenge of integration into existing PV manufacturing systems. Furthermore, the organic layers deposited on Si are thin, typically limited to <50 nm by the triplet diffusion length, resulting in <20 % photon absorption.[19] The second strategy is the harvesting of dark triplet excitons generated via SF by an inorganic semiconductor quantum dot emitter (QD) material, followed by the emission of light from the QD and its optical coupling into the PV module, as illustrated in Figure 1.[8,9,20,21] This approach converts the exciton multiplication process into a photon multiplication process. Since the output of such a singlet fission photon multiplication film (SF-PMF) is photons, they can be directed towards the PV cell without any change in cell design. Direct contact between the SF-PMF and the PV cell is not required and the SF-PMF could be integrated into the PV module in a number of different ways.

However, to achieve such as SF-PMF a number of materials and film processing challenges must be overcome. Specifically, a composite film of the organic SF material and QD emitters, with sufficient thickness is required (typically >1 μm) so as to absorb nearly 100% of the photons above twice the bandgap of Si. Simultaneously, the absorption of photons below twice the bandgap of Si must be minimised as ideally these wavelengths would pass unimpeded to the underlying Si cell. This means that the number of QDs in the film must be minimised. They must also be well dispersed, both to avoid aggregation-induced quenching and also to ensure that QDs are always within the diffusion length of triplet excitons generated in the organic phase. Furthermore, SF must occur efficiently in the organic phase and the interface between the organic and QD must also allow for efficient triplet harvesting. This leads to the idealised picture of a SF-PMF as illustrated in Figure 1a, where individual QDs are blended into the organic with the spacing between them set by the triplet diffusion length.

In practise, achieving such a morphology of well dispersed QDs in the organic, while preserving efficient SF in the organic phase, followed by efficient triplet diffusion to the QD, as well as efficient triplet energy transfer (TET) into the QD is extremely challenging. This is because the components of organic semiconductor-quantum dot blends have a strong tendency to aggregate and phase separate during processing, due to a mismatch of their size, shape and surface energies,[10–12] with detrimental effects on device performance[22–25]. It is important to note that this problem is not unique to SF-PMFs but applies broadly to organic-QD blends for photovoltaics, photodetectors and light-emitting diodes (LEDs). In each of these areas the underlying combination of organic and QD holds great promise, but blend morphology has long compromised device performance.

Here, we demonstrate a QD surface engineering approach using an electronically active, highly soluble semiconductor ligand that is matched to the organic host material, which allows us to direct the self-assembly of these blends, achieving optimal morphology while preserving electronic properties of both components as well as the interface between them. Our results open a clear route for SF-PMFs to improve the efficiency of Si PVs, as well as providing design rules to fabricate bulk organic-QD blends for a variety of applications.[26–32]

As SF host material we chose 5,12-bis((triisopropylsilyl)ethynyl)tetracene (TIPS-Tc), Figure 1a, a highly soluble, solution-processable molecular semiconductor shown to have a SF triplet yield of 130-180% in polycrystalline films.[33] As TIPS-Tc has a $T_1$ energy in the range 1.1-1.2 eV, and triplet transfer to the quantum dot should be exothermic, PbS quantum dots with an exciton peak absorption at 1.08 eV were used as the IR-emitting acceptor (Figure S1). The photoluminescence quantum efficiency (PLQE) of the as-synthesised PbS quantum dots ligated with oleic acid (PbS-OA) was 31% in toluene. We recently demonstrated that a tetracene-based ligand is necessary to achieve efficient triplet transfer in solution from TIPS-Tc to PbS quantum dots.[34] The aliphatic organic ligands [e.g. oleic acid (OA)] that are bound to the surfaces of QDs following their synthesis inhibit the transfer of triplet excitons to the QD in solution. As we develop below, these aliphatic organic ligands, which have been widely used in organic-QD blends to date, lead to phase segregation and QD aggregation upon film formation. Hence, PbS-OA quantum dots were ligand exchanged with the 'active-ligand' - 6,11-bis((triisopropylsilyl)ethynyl)tetracene-2-carboxylic acid (TET-CA), Figure 1a, to obtain TET-CA-ligated PbS quantum dots (PbS-TET-CA). In doing so the quantum dot PLQE dropped slightly to 24% in toluene. The ligand exchange from PbS-OA to PbS-TET-CA was quantified by combining small-angle X-ray and neutron scattering measurements (SAXS and SANS, respectively). SAXS was employed to measure the PbS core radius and dispersity, with SANS providing insight into the changes in the quantum dot ligand envelope.[35] Figure 2a shows SANS data and associated fits, which indicate that the ligand shell neutron scattering length density increases upon ligand exchange in a way that is consistent with a TET-CA ligand density of approximately 0.6 ± 0.1 ligands/nm$^2$, which was confirmed via measurements of optical absorption. Further, the exchanged quantum dot ligand shell may be described by $\rho_{shell} = \phi_{TET-CA}\rho_{TET-CA} + \phi_{OA}\rho_{OA} + \phi_{solvent}\rho_{solvent}$ (where $\rho$ = scattering length density and $\phi$ = volume fraction), with the TET-CA ligand scattering length density consistent with values of $\phi_{TET-CA}$ = 0.35, $\phi_{OA}$ = 0.42 and $\phi_{solvent}$ = 0.23, leading to the conclusion that significant TET-CA functionalisation of the PbS quantum dot has been achieved, but that residual OA is also present. (For full details of solution SAXS/SANS see Supporting Information Sections 2.1 and 2.2).

Bulk SF-PMFs were fabricated via blade coating solutions of TIPS-Tc (100 mg/mL) with either PbS-OA or PbS–TET-CA QDs (50 mg/mL) from toluene, allowing for film thickness ranging from 5 to 10 μms to be readily achieved. Solution casting TIPS-Tc by itself from low-volatility solvents readily generates highly crystalline, spherulitic type morphologies. We find that blends of TIPS-Tc with the unmodified QDs (PbS-OA) form morphologies in which the QDs are highly aggregated, while QDs modified with the TET-CA ligand yield much more dispersed morphologies. Figures 2b-d show grazing incidence small-angle X-ray scattering (GISAXS) from the QD:TIPS-Tc films, with clear structure factors between 0.05 and 0.35 Å$^{-1}$, representing colloidal crystallisation of aggregated quantum dots. Fits capturing the most significant features of the 1D radially integrated scattering data, were obtained for both PbS-OA:TIPS-Tc and PbS-TET-CA:TIPS-Tc films using a colloidal crystal model (see Supporting Information for full fit parameters). The corresponding fit parameters for the PbS-OA:TIPS-Tc film, lattice constant of 90.1 ± 3.9 Å and lattice distortion factor of 0.08, indicate the formation of highly ordered QD aggregates. In contrast, the PbS–TET-CA blends show much weaker colloidal ordering, with fit parameters; lattice constant of 76.8 ± 9.6 Å representing a decrease in the apparent thickness of TET-CA containing ligand shell, and a greatly increased lattice distortion factor of 0.25 indicating a significantly enhanced contact between QDs and the singlet-fission host. The insets of Figure 2b-c shown polarised optical microscopy (POM) images, indicating the presence of large radially orientated crystalline domains of the TIPS-Tc host, for both PbS-TET-CA:TIPS-Tc and PbS-OA:TIPS-Tc blend films respectively. However, the nucleation density is much lower for the PbS-OA:TIPS-Tc films, suggesting that the PbS-TET-CA quantum dots are involved in the TIPS-Tc crystallisation process.

The GISAXS data quantitatively show that PbS-OA QDs form highly ordered aggregate structures within TIPS-Tc, whilst PbS-TET-CA QDs are more randomly distributed within TIPS-Tc. This is further confirmed via TEM imaging shown in Figures 2e & f and S2 & S3. Thus, the conventional aliphatic OA ligand which has unfavourable interactions with the TIPS-Tc leads to self-assembly processes during film formation that gives rise to phase segregation and QD aggregation, as illustrated in Figure 2g. In contrast, the favourable interaction between the active TET-CA ligand and the bulk TIPS-Tc matrix allows for a directed self-assembly process, where phase segregation and QD aggregation are arrested, as illustrated in Figure 2h.

We now turn to characterising the optoelectronic properties of these films. The absorption and photoluminescence of the PbS-TET-CA:TIPS-Tc films are shown in Figure 3a. All films studied herein have an absorbance greater than 1.5 (>95 % absorption) at the TIPS-Tc absorption peak (545 nm), with PbS QDs absorbance an order of magnitude lower across the visible region (Figure 3a). Figure 3b displays the IR PL excitation spectra of a PbS-OA:TIPS-Tc film and qualitatively shows that there is relatively low triplet transfer from the TIPS-Tc to the PbS-OA QDs. The reduction in IR PL at excitation wavelengths where the TIPS-Tc is absorbing indicates that the TIPS-Tc is "shadowing" the QDs resulting in lower QD emission. In contrast, excitation spectra of PbS-TET-CA:TIPS-Tc films show high levels of energy transfer from TIPS-Tc to the PbS QDs, which we show below to arise from triplet excitons.

To quantitatively evaluate the photon multiplication performance in the films, we measure the PLQE when exciting the SF host material TIPS-Tc (at 515 nm) and compare to direct excitation of the PbS-TET-CA QDs (at 658 nm). The PLQE increases from (15.4 ± 1.0) % (658 nm exciting only QDs in the blend) to (24.5 ± 1.0) % (515 nm, exciting both components of the blend). This enhancement of (59 ± 12) % suggests efficient SF followed by triplet energy transfer (TET) to the emissive QDs. Using the relative absorption in the PbS-TET-CA quantum dots and the SF host, an exciton multiplication factor of $\eta_{EMF}$ = (186 ± 18) % can be estimated from Equation S1, $\eta_{EMF}$ is also given by the product of the singlet fission yield $\eta_{SF}$ and the triplet transfer efficiency $\eta_{TET}$. $\eta_{EMF}$ serves as a metric of blend performance as it captures both the retention of SF properties of the SF host material, the quality of morphology via the efficient diffusion of triplets to the QDs and quality of the interface via the transfer of the triplets into the QD. In contrast to the PbS-TET-CA:TIPS-Tc films, films of PbS-OA:TIPS-Tc show a drop in PLQE when the SF host is excited, from (17.2 ± 1) % (658 nm exciting only QDs in the blend) to (3.8 ± 1) % (515 nm, exciting both components of the blend), as shown in Table 1.

To verify that the PLQE enhancement originates from SF and triplet transfer we perform magnetic-field-dependent PL measurements (Figure 3c). We observe an initial decrease in TIPS-Tc PL at low magnetic fields (<150mT) followed by an increase at higher magnetic fields. This behaviour is typical for SF materials.[8,34] The quantum dot PL shows the opposite trend, demonstrating that the QD PL arises from triplet energy transfer to the QDs.

To further characterise the PL enhancement and gain a mechanistic understanding of the triplet transfer to the QD, we used ns-resolved transient absorption (ns-TA) and near-infrared (NIR) time-correlated single photon counting (TCSPC). The spectrally narrow TIPS-Tc triplet photoinduced absorption (PIA) features at 850 and 970 nm readily lend themselves to the extraction of the triplet population dynamics in the blend films (Figure S5 – S8). The PIA kinetics (965-980 nm) in pristine TIPS-Tc films and PbS-TET-CA:TIPS-Tc films were fitted to extract the intrinsic triplet decay rate $k_1$ and the triplet transfer rate $k_{TET1}$. A comparison of the TIPS-Tc triplet lifetime in films of TIPS-Tc, PbS-OA:TIPS-Tc and PbS-TET-CA:TIPS-Tc is shown in Figure 4a. Based on the increase in monomolecular TIPS-Tc triplet decay rate a significant triplet quenching of (97 ± 11) % is estimated for the film containing PbS-TET-CA quantum dots. Assuming that this triplet quenching is caused by triplet transfer to the QDs we estimate the singlet fission yield to be $\eta_{SF}$ = (192 ± 28) %, based on $\eta_{TET}$ = (97 ± 11) % and the exciton multiplication factor $\eta_{EMF}$ = (186 ± 18) % calculated above. The high SF yield is in line with previous reports and indicates that the process is not influenced by the doping of TIPS-Tc with PbS-TET-CA QDs (see also Figure S5).[33]

As shown in Figure 4b (also seen in Figures S6, S7 and S9), it is apparent that under direct excitation (650 nm) the peak of the ground-state bleach (GSB) signal for the PbS-OA QDs in the blend shifts to lower energies by roughly 60 meV within 20 ns, whereas the PbS-TET-CA QD in the blend have a less than 10 meV shift over their entire excited state lifetime. Comparing the rate of QD decay with the rate of aggregation-assisted hopping to low-energy quantum dot sites (observed by the dynamic red shifting of the QD GSB) indicates that (90 ± 20) % of PbS-OA excitations find a lower-energy QD site. On the other hand, only (17 ± 2) % of PbS-TET-CA excitations find lower-energy QD sites within the SF host (see Supporting Information Section 7.1 and Figures S9 & S10 for details).[12,36] The enhanced aggregation-assisted hopping in the PbS-OA:TIPS-Tc films is consistent with the morphological insights afforded from GISAXS measurements discussed above.

Time-correlated single photon counting was employed to measure the IR transient PL under excitation of the QDs (650 nm) and the SF host (535 nm) in both PbS-TET-CA:TIPS-Tc and PbS-OA:TIPS-Tc films (Figures 4c and S12-14). Figure 4c compares the normalised PL kinetics of a PbS-TET-CA:TIPS-Tc film under different excitation wavelengths, and shows higher levels of delayed QD PL when the SF host is excited (535 nm) relative to the intrinsic quantum dot dynamics (650 nm). To extract the triplet transfer dynamics, the QD intrinsic decay (650 nm excitation) was deconvolved from the QD decay with triplet transfer (535 nm excitation) using a fast Fourier transform (FFT) (see Supporting Information Section 9.1).[21] The triplet transfer to the PbS QDs was found to occur on the μs timescale. To best model the transfer dynamics a three-state model was implemented (see Supporting Information Section 9.1-9.2 for details). Figure 4d summarises the transfer dynamics where the TIPS-Tc triplet is transferred to an intermediate state with rate $k_{TET1}$ (see above), followed by a second triplet transfer with rate $k_{TET2}$ to the PbS quantum dot. We tentatively assign the intermediate state as the TET-CA ligand triplet.[37]

The intrinsic quantum dot decay rate $k_{QD}$ can be determined from both ns-TA kinetics and the IR-TCSPC as well as extracted from the fit of the 3-state model with reasonable consistency (See supporting information for a detailed discussion). In Table 2 the rate constants used for the modelling are summarised, including the intrinsic quantum dot decay rate obtained from the fit ($k_{QD}$). The value for triplet mono-molecular decay, $k_1$, in TIPS-Tc is consistent with similar systems.[38–40]

In summary, we have demonstrated a route to overcoming the longstanding problems of phase-segregation and aggregation in solution processed organic-QD blends. From our spectroscopic and morphological characterisation, we can conclude that the role of the TET-CA ligand is two-fold; it enables efficient triplet transfer from the bulk SF host to the QDs and plays a key role in altering the surface chemistry of the QDs thereby achieving a directed self-assembly process which allows both optimal morphology while preserving electronic properties of both components as well as the interface between them. In the bulk SF-PMFs fabricated with PbS-TET-CA:TIPS-Tc, efficient SF ($\eta_{SF}$ = 192 ± 28 %) in the TIPS-Tc is followed by efficient triplet energy transfer to the QDs ($\eta_{TET}$ = 97 ± 11 %) giving rise to a ~190% exciton multiplication factor (out of a possible 200%). In comparison, films fabricated with conventional aliphatic ligands show strong QD aggregate formation and very poor exciton harvesting. The triplet energies of the SF host and active ligand used in this study mean that QDs with bandgaps (1.08eV) below that of Si (1.1eV) must be used to harvest the triplets, leading to emission of photons that cannot be absorbed by Si. The maximum PLQE of as synthesised QDs at these energies are also low, typically around 30%, whereas the PLQE of dots emitting at 1000nm, a wavelength that is readily absorbed in Si, are 80%. Hence, while we do not demonstrate the coupling of the SF-PMF directly to a Si PV in this manuscript, our results provide a clear proof of concept demonstration that establishes SF-PMF as a highly promising architecture to harness the SF process and provides clear roadmap for future work, where using higher triplet energy SF host materials and ligands should enable external quantum efficiencies (EQEs) above 100% in Si. More broadly, our results offer a general strategy to control morphology for a range of applications where organic-QD blends are desired, for instance photon upconversion systems based on organic-QD systems, which have so far been limited to solution phase or bilayers; or LEDs based on emitter

QDs within an organic host, which would benefit from this strategy to avoid aggregation induced quenching effects.

## Acknowledgements

The authors acknowledge funding through the Winton Programme for the Physics of Sustainability and the Engineering and Physical Sciences Research Council (UK). The authors acknowledge beamtime awarded at the ISIS Pulsed Neutron and Muon Source through experiment number RB1810513 (DOI: 10.5286/ISIS.E.RB1810513). VG acknowledges funding from the Swedish Research Council, Vetenskapsrådet 2018-00238. J. R. A. acknowledges Cambridge Commonwealth European and International Trust for financial support. J. X. acknowledges EPSRC Cambridge NanoDTC, EP/L015978/1 for financial support. JEA acknowledge the U.S. National Science Foundation (DMREF-1627428) for support of organic semiconductor synthesis. Z. Z. acknowledges funding from the European Union's Horizon 2020 research and innovation programme under the Marie Skłodowska-Curie Actions grant (No. 842271 – TRITON project). This project has received funding from the European Research Council (ERC) under the European Union's Horizon 2020 research and innovation programme (grant agreement number 758826).

Data availability: The data underlying this manuscript are available at [url to be added in proof].

No custom computer code or algorithm was used to generate the results presented herein.

Author contributions:

Supplementary information is available for this paper at doi ,

AR, NCG and RHF are founders of Cambridge Photon Technology, a company commercialising advanced solar cell technologies, of which SD and JW are employees. The other authors declare no competing non-financial interests.

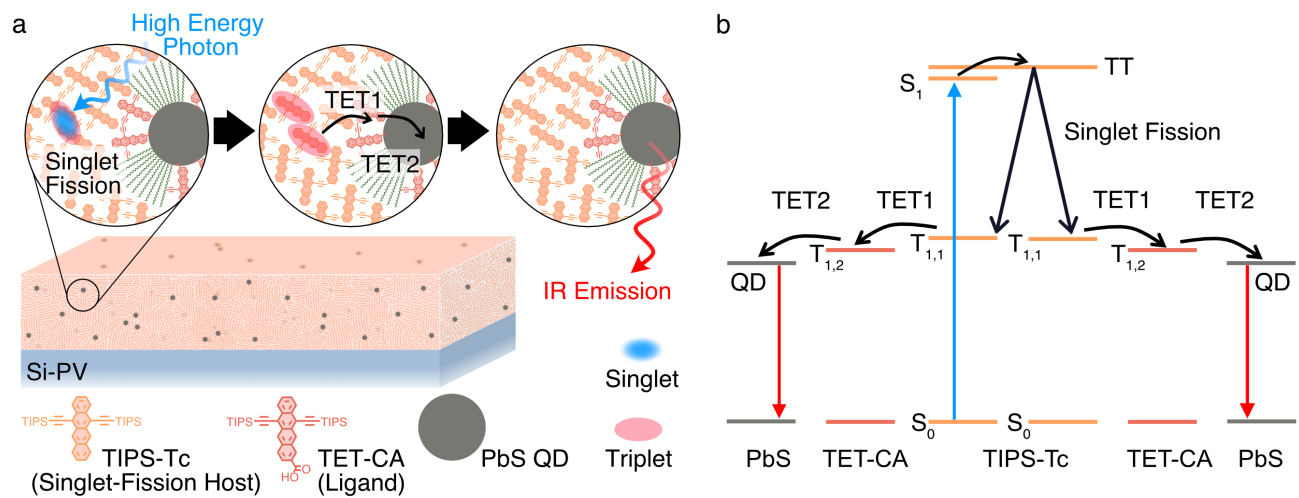

**Figure 1. Schematic illustration of a bulk singlet fission photon multiplier film integrated with a silicon-based photovoltaic device. (a)** Triplet acceptor PbS quantum dots are evenly dispersed within an optically dense, TIPS-Tc, singlet-fission host. **(b)** Energy diagram describing the processes involved in photon multiplication based on singlet fission. First, a high-energy photon is absorbed, followed by rapid singlet-fission in TIPS-Tc generating two triplet excitons ($T_{1,1}$). The triplet excitons are transferred via a TIPS-Tc carboxylic acid (TET-CA) ligand to PbS quantum dots which emit a photon when returning to the ground state.

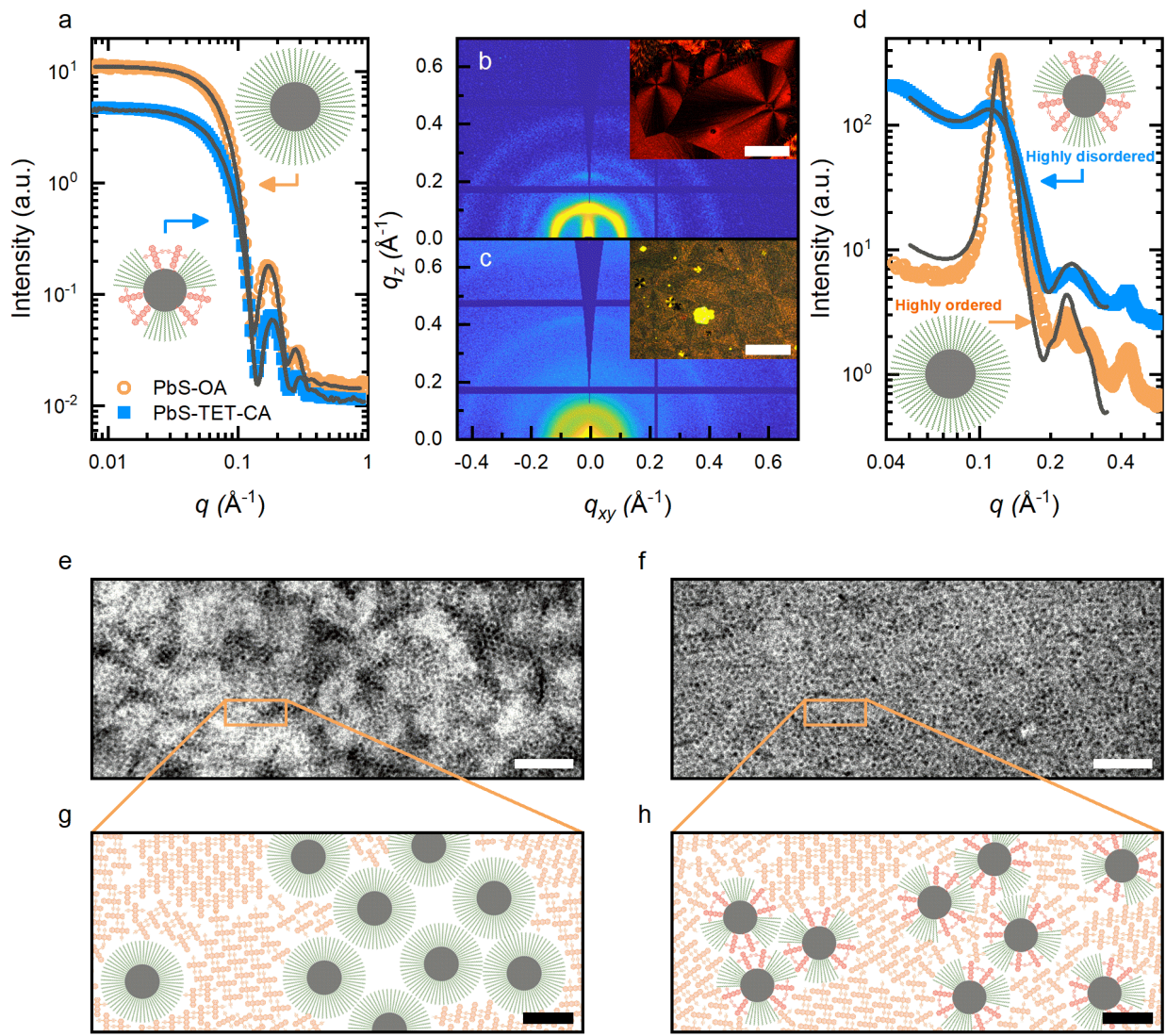

**Figure 2: Ligand dependence of the PbS quantum dots dispersion within the singlet fission host. (a)** SANS data from before and after ligand exchange, i.e. PbS-OA (orange open circles) and PbS-TET-CA (blue closed squares), following subtraction of appropriate backgrounds, with associated fits to a core-shell sphere*hard sphere model (black curves). Insets: schematic illustration of the population of ligands shifting from all OA (PbS-OA) to a mixture of OA and TET-CA (PbS-TET-CA). Two dimensional grazing incidence X-ray scattering data for PbS-OA:TIPS-TC **(b)** and PbS-TET-CA:TIPS-Tc films **(c)** inserts showing POM images (500 µm scale bar), with one dimensional radially integrated data shown **(d)**, with PbS-OA:TIPS-Tc (orange open circles), PbS-TET-CA:TIPS-Tc (blue closed squares) and associated fits to an FCC colloidal crystal model (black curves). TEM (50 nm scale bar) for PbS-OA:TIPS-Tc **(e)** showing large aggregates (dark regions) within the SF host (lighter regions) and PbS-TET-CA **(f)** showing a significantly more homogenous quantum dot dispersion within the TIPS-Tc host. Illustration (5 nm scale bar) of the bulk SF-PM structures for the highly ordered parking of the PbS-OA quantum dots **(g)** and the highly disordered dispersion of PbS-TET-CA quantum dots **(h)** within the TIPS-Tc.

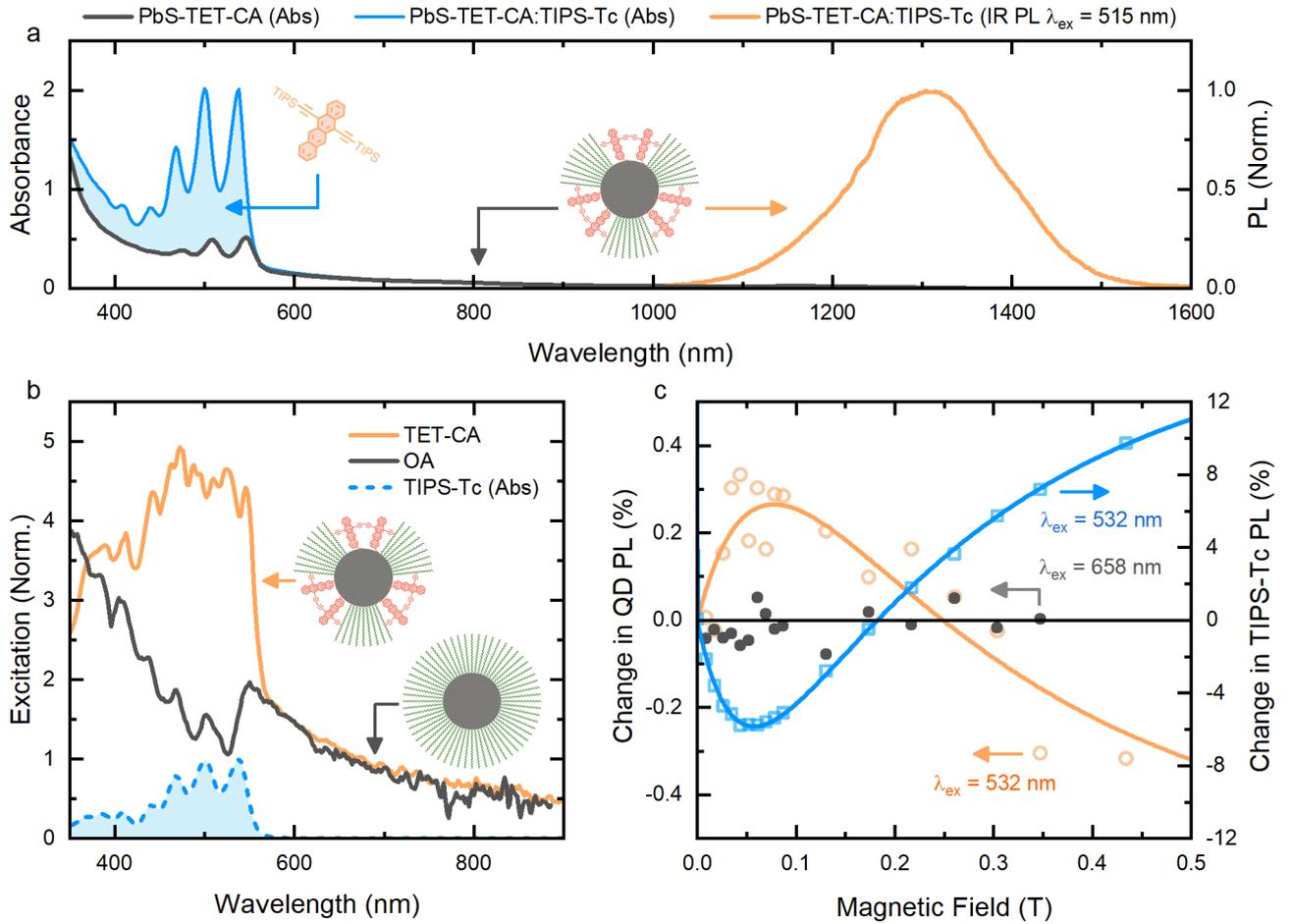

**Figure 3: Absorbance and steady-state IR PL resulting from triplet harvesting in a film of TIPS-Tc and PbS-TET-CA quantum dots. (a)** Absorbance (blue curve) and normalised IR PL (orange curve) of a PbS-TET-CA:TIPS-Tc thin film. For comparison the absorbance of PbS-TET-CA quantum dots in toluene (black curve), the difference highlighting the TIPS-Tc absorption (blue area). **(b)** QD IR PL excitation spectra of PbS-OA:TIPS-Tc (dark grey curve) and PbS-TET-CA:TIPS-Tc films (orange curve), normalised to the average value between 650-700 nm. PbS QD emission was collected in the wavelength range 1300 ± 20 nm. Also plotted is the normalised absorbance spectrum of TIPS-Tc (dashed blue curve). **(c)** Percentage change in PL from the PbS-TET-CA QDs (orange and grey circles) and TIPS-Tc (blue squares) in a PbS-TET-CA:TIPS-Tc film as a function of external magnetic field. The film was excited at either 532 nm (absorbed by SF and QD components) or 658 nm (selective excitation of the QD). Solid lines illustrate guides for the eye of both the QD (orange curve) and TIPS-Tc (blue curve) PL change under 532 nm excitation.

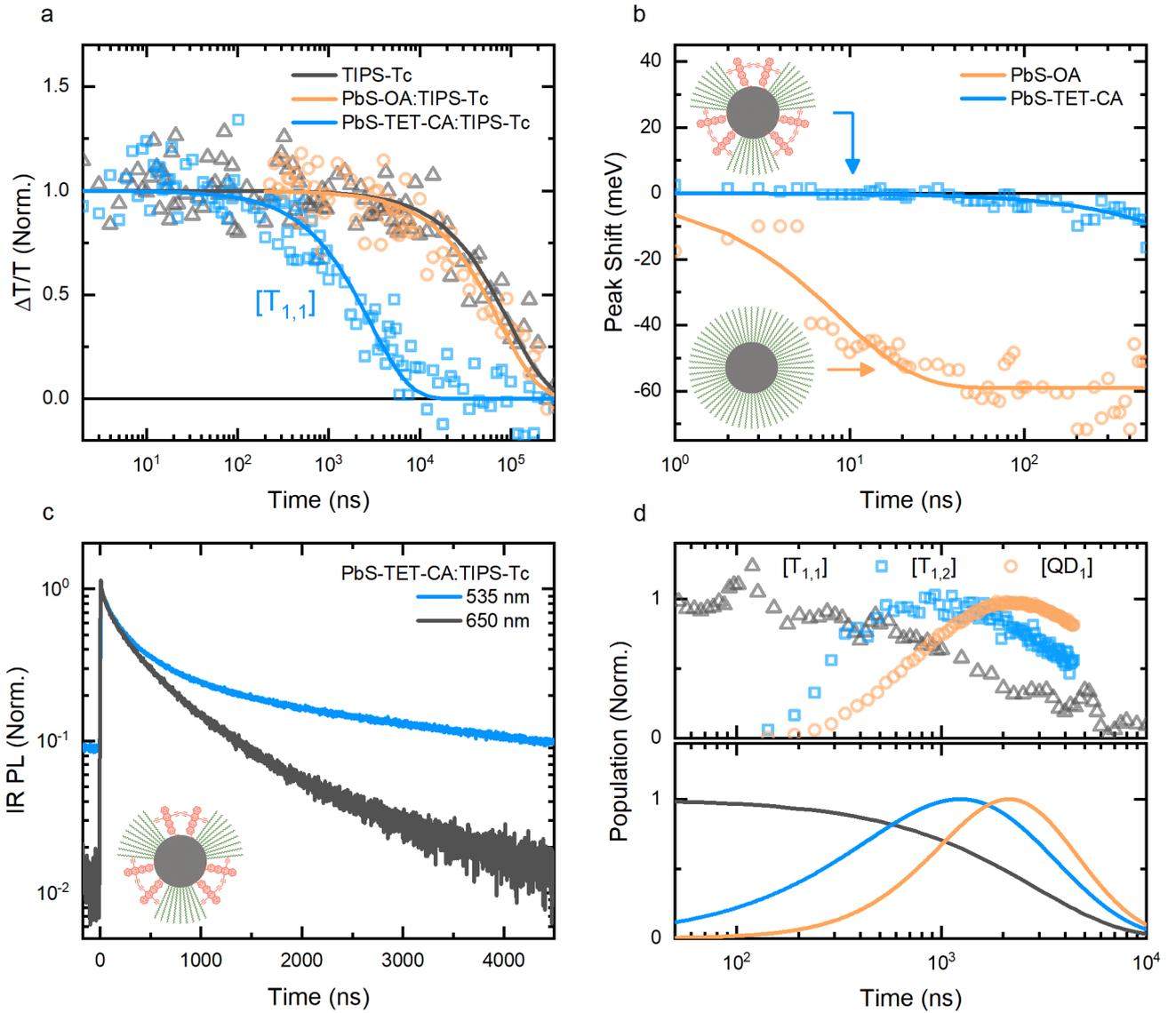

**Figure 4: Time-resolved triplet transfer and quantum dot excited state dynamics. (a)** Normalised transient absorption kinetics at the TIPS-Tc triplet PIA (965-980 nm) for films of TIPS-Tc, either pristine (light grey triangles), with PbS-OA (light orange circles) or with PbS-TET-CA quantum dots (light blue squares), excited at 515 nm with ~15 μJ/cm$^2$, with mono-exponential fits (darker curves). **(b)** The shift in the peak position of the PbS quantum dot's GSB, in films of PbS-OA:TIPS-Tc (light orange circles) and PbS-TET-CA:TIPS-Tc (light blue squares) after excitation with 658 nm at ~15 μJ/cm$^2$, with mono-exponential fits (darker curves). **(c)** Normalised transient IR PL for a film of PbS-TET-CA:TIPS-Tc under 535 nm (blue curve) and 650 nm excitation (black curve), at ~0.015 and 0.010 μJ/cm$^2$ respectively. **(d)** The decay of the TIPS-Tc triplets [$T_{1,1}$] (black) transferring to TET-CA triplets [$T_{1,2}$] (blue) and then finally leading to emission from the excited PbS quantum dot state [$QD_1$] (orange). **d, top)** [$T_{1,1}$] determined from nsTA, while the [$T_{1,2}$] and [$QD_1$] populations were determined from transient IR PL. **d, bottom)** Associated fits to the dynamics in the three-state kinetic scheme.

| Quantum Dot | PLQE ($\lambda_{ex}$ = 515 nm) | PLQE ($\lambda_{ex}$ = 658 nm) | $\eta_{EMF}$ |
|---|---|---|---|
| PbS-OA | (3.8 ± 1) % | (17.2 ± 1) % | (-4 ± 8) % |
| PbS-TET-CA | (24.5 ± 1.0) % | (15.4 ± 1.0) % | (186 ± 18) % |

**Table 1: Photoluminescence performance for thin films of PbS quantum dots in a TIPS-Tc SF host.** The PLQE of the PbS QD emission in films of PbS-OA:TIPS-Tc and PbS-TET-CA:TIPS-Tc under excitation at either 515 nm (absorbed by both SF and QD components) or 658 nm (absorbed by QD only). Based on these PLQE values and the relative absorption of the films components the exciton multiplication factor, $\eta_{EMF}$, is calculated.

| $k_1$ (1/μs) | $k_{TET1}$ (1/μs) | $k_{TET2}$ (1/μs) | $k_{QD}$ (1/μs) |
|---|---|---|---|
| (1.05 ± 0.10)x10$^{-2}$ | 0.34 ± 0.03 | 1.6 ± 0.1 | 1.40 ± 0.2 |

**Table 2: Kinetic parameters for a thin film of TET-CA capped PbS quantum dots in a TIPS-Tc SF host.** The monomolecular TIPS-Tc triplet decay rate $k_1$ and TIPS-Tc to PbS-TET-CA triplet transfer rate $k_{TET1}$ are measured by fitting to the ns-TA. The TET-CA to PbS quantum dot triplet transfer rate $k_{TET2}$ and PbS quantum dot intrinsic decay rate $k_{QD}$ are extracted by global fitting of the PbS-TET-CA:TIPS-TC IR PL transient dynamics.

# Supplementary Information: Ligand Directed Self-Assembly of Bulk Organic-Semiconductor/Quantum-Dot Blend Films Enables Near Quantitative Harvesting of Triplet Excitons

Jesse Allardice[a*], Victor Gray[a,b*], Simon Dowland[c*], Daniel T. W. Tooland[d*], Michael P. Weir[e*], James Xiao[a], Zhilong Zhang[a], Jurjen F. Winkel[c], Anthony J. Petty II[f], John Anthony[f], Richard Friend[a], Anthony J. Ryan[d], Richard A. L. Jones[e], Neil C. Greenham[a] and Akshay Rao[a†]

**Contacts**

[a]Cavendish Laboratory, University of Cambridge, J. J. Thomson Avenue, Cambridge, CB3 0HE, UK

[b]Department of Chemistry, Ångström Laboratory, Uppsala University, Box 532, SE-751 20 Uppsala, Sweden

[c] Cambridge Photon Technology, J. J. Thomson Avenue, Cambridge, CB3 0HE, UK

[d]Department of Chemistry, The University of Sheffield, Sheffield S3 7HF, United Kingdom

[e]Department of Physics and Astronomy, The University of Sheffield, Sheffield S3 7RH, United Kingdom

[f]Center for Applied Energy Research, University of Kentucky, Research Park Dr., Lexington KY 40511, USA

*: Authors contributing equally to this work.

†: Corresponding Author.

# 1 Contents

## 2 Methods

### 2.1 Small-Angle Neutron Scattering

SANS was carried out on the SANS2D[1] small-angle diffractometer at the ISIS Pulsed Neutron Source (STFC Rutherford Appleton Laboratory, Didcot, U.K.).[2] Samples before (PbS-OA) and after ligand exchange (PbS-TET-CA) were prepared in deuterated toluene, providing the necessary contrast, and were contained in 2 mm path length quartz cells (Hellma GmbH). In the following, the magnitude of the scattering vector is defined as $q = \frac{4\pi sin\theta}{\lambda}$ where $2\theta$ is the angle between the incident and scattered X-ray or neutron of wavelength. A simultaneous $q$-range of 0.006 – 1.2 Å$^{-1}$ was achieved utilizing an incident wavelength range of 1.65 – 16.5 Å and employing an instrument set up of L1 = L2 = 4 m, where L1 and L2 are the pre and post sample flightpaths respectively, with the rear detector offset vertically 75 mm and horizontally 100 mm. The beam diameter was collimated to 12 mm at the sample. For all collected data, each raw scattering data set was corrected for the detector efficiencies, sample transmission and background scattering and converted to scattering cross-section data (∂Σ/∂Ω vs. $q$) using the instrument-specific software.[3] These data were placed on an absolute scale (cm$^{-1}$) using the scattering from a standard sample (a solid blend of hydrogenous and perdeuterated polystyrene) in accordance with established procedures.[4]

Appropriate backgrounds were subtracted from each sample, namely oleic acid in d-toluene for PbS-OA[5] and TIPS-tetracene in d-toluene for PbS-TET-CA, to approximate the scattering from residual TET-CA in solution following ligand exchange. Fitting was performed using the *SasView* software package.[6] The data were fitted to a core-shell sphere model with a hard-sphere structure factor. The core radius of 16.3 Å and lognormal polydispersity of 0.1 were used as constraints to the SANS fitting, detailed procedures for the characterisation of this PbS-OA sample set is described elsewhere.[5] The main output of the model was therefore a scattering length density, thickness, and polydispersity of the ligand shell. In all cases, the component scattering length densities are taken as $\rho_{OA}$ = -0.24 (packed OA tails), $\rho_{TET-CA}$ = 0.9 and $\rho_{d-toluene}$ = 5.68 × 10$^{-6}$ Å$^{-2}$. In the case of PbS-OA, the shell SLD is simply given by $\rho_{shell} = \phi_{OA}\rho_{OA} + \phi_{solvent}\rho_{solvent}$ allowing simple calculation of the respective volume fractions within the shell of oleic acid and solvent as $\phi_{OA}$ = 0.83 and $\phi_{solvent}$ = 0.17. In the case of PbS-TET-CA, a ligand density on the order of 0.6 ± 0.1 ligands/nm$^2$ is typically measured by absorption. The exchanged ligand shell is therefore described by $\rho_{shell} = \phi_{TET-CA}\rho_{TET-CA} + \phi_{OA}\rho_{OA} + \phi_{solvent}\rho_{solvent}$. For a TET-CA ligand density of 0.6 ± 0.1 ligands/nm$^2$, this is consistent with values of $\phi_{TET-CA}$ = 0.35, $\phi_{OA}$ = 0.42 and $\phi_{solvent}$ = 0.23, leading to the conclusion that significant TET-CA functionalisation of the PbS has been achieved, but that residual OA is also present.

## 2.2 Grazing Incidence X-ray Scattering

Grazing incidence X-ray scattering measurements were carried out on a Xeuss 2.0 instrument equipped with an Excillum MetalJet liquid gallium X-ray source. For films prepared on glass, X-rays were collected for 900 s using collimating slits of 0.25 × 0.3 mm ("high resolution" mode). Alignment was performed via three iterative height (z) and rocking curve (Ω) scans, with the final grazing incidence angle set to Ω = 0.1º. Scattering patterns were recorded on a vertically-offset Pilatus 1M detector with a sample to detector distance of 559 mm, calibrated using a silver behenate standard to achieve a $q$-range of 0.045 – 1.2 Å$^{-1}$. Data reduction was performed using the instrument-specific *Foxtrot* software. Two-dimensional scattering data was reduced to one-dimensional via radial integration, which was performed with a mask to remove contributions from "hot pixels", substrate horizon and the reflected beam. Fitting was performed using the *SasView* software package.[6] The data were fitted to the FCC paracrystal model.

|  | TIPS-Tc:PbS-OA | TIPS-Tc:PbS-TET-CA |
|---|---|---|
| scale | 0.60943 | 0.54412 |
| background | 0.15081 | 3.034 |
| Lattice constant | 89.719 | 76.76 |
| Lattice order parameter | 0.085943 | 0.2499 |
| PbS core radius (Å) | 22.34 | 22.34 |
| PbS core log normal polydispersity (Å) | 0.079987 | 0.075 |
| PbS X-ray sld(Ga) (10$^{-6}$ Å$^{-2}$) | 50.7 | 50.7 |
| TIPS-Tc X-ray sld (10$^{-6}$ Å$^{-2}$) | 10.2 | 10.2 |

*Table S1: Fit parameters for TIPS-Tc:PbS-OA and TIPS-Tc:PbS-TET-CA scattering data.*

## 2.3 TEM

Transmission electron microscopy to investigate film morphology was performed using an FEI Tecnai F20 at 200 kV accelerating voltage. Film samples at 50 mg mL$^{-1}$ QDs and 100 mg mL$^{-1}$ TIPS-Tc were transferred onto 200-mesh Cu grids (Agar AGS160).

## 2.4 Ligand Exchange

Synthesis of PbS QDs was carried out following the procedure by Hines and Scholes with modifications[7, 8] In summary, PbO (0.45 g), oleic acid (8 g) and 1-octadecene (10 g) were degassed in a three-necked flask at 110 °C for 2 h. The temperature was then reduced to 95 °C. Under nitrogen, a solution of bis(trimethylsilyl)sulphide (210 µL) in 1-octadecene (5 mL) was rapidly injected into the lead precursor solution. After cooling naturally to room temperature the PbS QDs were washed 4 times by precipitation/re-dispersion with acetone and hexane. The purified QDs were stored in a nitrogen filled glovebox at high concentration (>40 mg mL$^{-1}$ / >100 µM) until use. Ligand exchange was carried out under nitrogen. The QDs in toluene were diluted to 8 mg mL$^{-1}$ in a Toluene/THF mixture of 4:1. The ligand in 100 mg mL$^{-1}$ THF solution was added to the QD solution, keeping a ligand to QD mass ratio of 1:1.

Ligand coverage was estimated from UV/Vis absorption using the molar absorption coefficient of the TET-CA ligand as 25500 M$^{-1}$cm$^{-1}$ at the peak absorption in toluene. The ligand area was used assuming the QD as a sphere with a diameter estimated from the empiric formula in Moreels *et al.*[9]

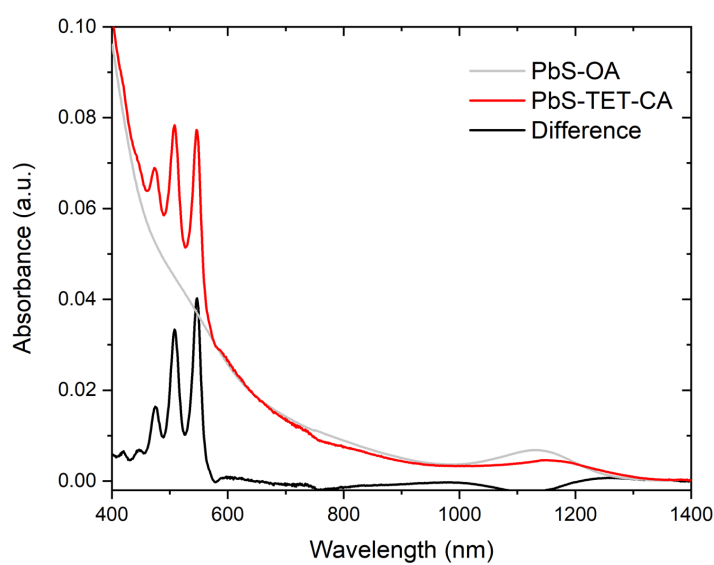

*Figure S1: Absorption of PbS-OA, PbS-TET-CA after ligand exchange and difference spectra used to calculate the ligand coverage.*

## 2.5 Steady-State Absorption

A Shimadzu UV-3600Plus spectrometer was used to measure the absorbance spectra of the solutions and films.

## 2.6 PLQE and Excitation Scan

The integrating sphere and PLQE measurement procedure has been described previously.[10,11] In summary, an integrating sphere with a Spectralon-coated interior (Newport 819C-SL-5.3) was used. 515 nm (2.9 x 10$^{15}$ photons s$^{-1}$cm$^{-2}$ at the sample) and 658 nm (1.8 x 10$^{16}$ photons s$^{-1}$cm$^{-2}$ at the sample) laser diodes (Thorlabs) with a beam diameter at the sample of 3 mm was used as the excitation source. Light from the sphere was coupled into an Andor Kymera 328i Spectrograph equipped with an InGaAs detector (Andor, iDus InGaAs 490). From the reproducibility of consecutive measurements, we estimate the absolute uncertainty on a PLQE value at 1 unit %.

Excitation scans were recorded on an Edinburgh Instruments FLS 980 Fluorimeter using an InGaAs detector for recording the IR emission (1300 ± 20 nm).

## 2.7 Transient Absorption

The short time (fs-ns) transient absorption setup has been described previously.[11] In summary, a Light Conversion PHAROS laser system with 400 µJ per pulse at 1030 nm with a repetition rate of 38 kHz was used. The output is divided, one part is focused onto a 4 mm YAG substrate to produce the continuum probe beam from 520 to 950 nm. The second part of the PHAROS output is lead into a narrow band optical parametric oscillator system (ORPHEUS-LYRA, Light conversion) outputting the pump beam. The probe pulse is delayed up to 2 ns with a mechanical delay-stage (Newport). A mechanical chopper (Thorlabs) is used to create an on-off pump-probe pulse series. The pump size on the sample is approximately 0.065 mm$^2$ and the probe about 0.015 mm$^2$. A silicon line scan camera (JAI SW-2000M-CL-80) fitted onto a visible spectrograph (Andor Solis, Shamrock) is used to record the transmitted probe light.

The longtime (ns-µs) transient absorption setup has also been described previously.[11] In short, the pump-probe setup consists of a probe from a LEUKOS Disco 1 UV supercontinuum laser (STM-1-UV, 1 kHz) and a pump generated in a TOPAS optical amplifier, pumped with the output from a Spectra-Physics Solstice Ace Ti:Sapphire amplifier (1 kHz). The probe beam is split into a reference and probe and both are focused onto the sample. A pair of line image sensors (Hamamatsu, G11608) mounted on a spectrograph (Andor Solis, Shamrock SR-303i) is used to detect the signal, using a custom-built board from Stresing Entwickslungsburo to read out the signal.

## 2.8 IR TCSPC

Samples were excited with a pulsed supercontinuum laser (Fianum Whitelase SC-400-4, 6 ps pulse length) at 0.2 MHz repetition rate. The pump wavelength set to either 535 nm or 650 nm (full-width at half-maximum 10 nm) with dielectric filters (Thorlabs). Pump scatter from the laser excitation within the photoluminescence path to the detector was filtered-out with an absorptive 900 nm long-pass filter (Thorlabs). The infrared photoluminescence was focused and detected by a single-photon avalanche photodiode based on InGaAs/InP (MPD-InGaAs-SPAD).

## 2.9 Steady-State and Magnetic Dependent PL

Similar to previously described temperature and current-controlled laser diodes (Thorlabs) were used to generate stable 532 nm and 658 nm laser beams.[11] The incident beam was attenuated as desired and focused onto the sample while PL from the sample was collected and focused into an Andor Kymera 328i Spectrometer and spectra recorded using either a Si-CCD (Andor iDus 420) for the visible region or an InGaAs detector (Andor, Dus InGaAs 490) for the NIR region. For magnetic field dependent PL measurements, an electromagnet was placed such that the sample was located within the poles of the electromagnet. As described previously,[12] the electromagnet was driven to achieve varying magnetic field strengths by using a Keithley 2400 variable voltage source connected to a current amplifier. The magnetic field between the poles (at the sample position) was calibrated to the applied voltage by a Gauss-meter. When measuring the PL from the PbS QDs (near IR region), both an RG1000 (Schott) and PL950 (Thorlabs) long pass filters were placed in front of the entrance to the spectrometer. These removed laser scatter and higher order peaks from the grating. After averaging over multiple sweeps of the magnetic field and integration of the spectra, the percentage change relative to the spectrum under zero applied field strength was calculated.

## 3 TEM

Figures S2 and S3 show TEM images of PbS-OA:TIPS-Tc and PbS-TET-CA:TIPS-Tc films. The PbS-OA and PbS-TET-CA QDs displayed similar size distributions as imaged within TIPS-Tc films. PbS-OA showed significant aggregation whilst PbS-TET-CA showed well-dispersed QDs despite high QD loadings. The more favourable interaction of TET-CA ligands with the TIPS-Tc compared to oleic acid prevented the phase segregation behaviour observed in the PbS-OA films.

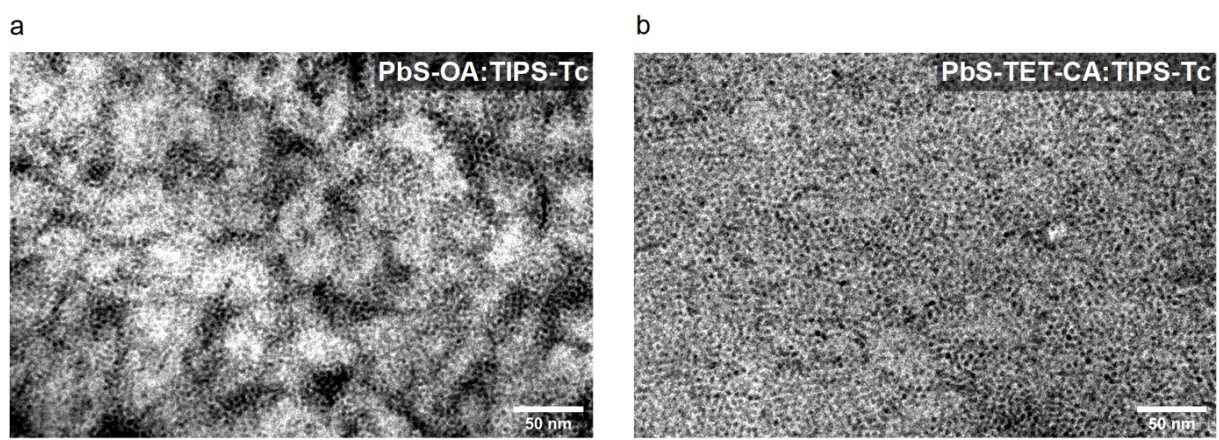

*Figure S2: TEM at the 50 nm scale, a) PbS-OA:TIPS-Tc b) PbS-TET-CA:TIPS-Tc films. The PbS-OA:TIPS-Tc films show clear phase separation, with large aggregates of PbS-OA QDs (dark dots) within the SF-host (lighter regions). While the PbS-TET-CA:TIPS-Tc films are significantly more homogeneously dispersed throughout the TIPS-Tc SF-host material.*

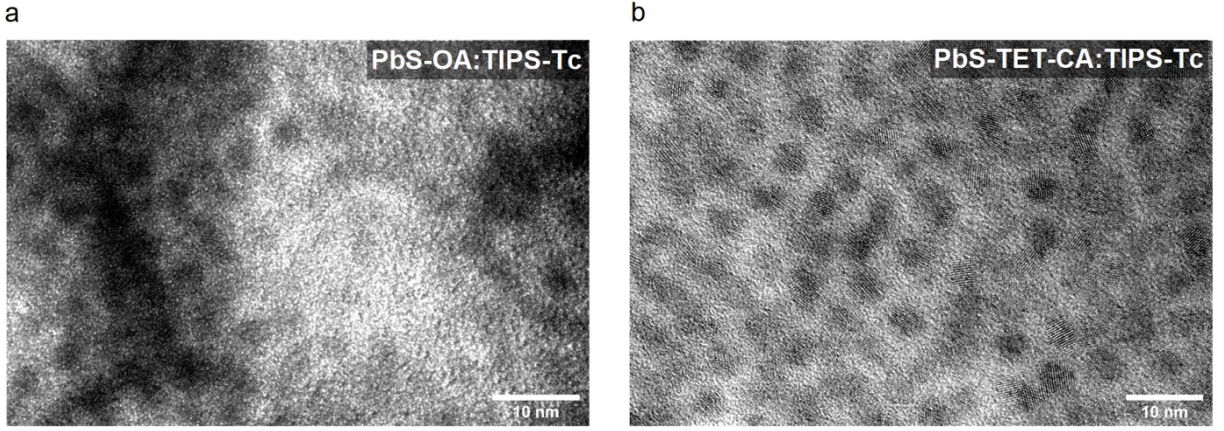

*Figure S3: TEM at the 10 nm scale, a) PbS-OA:TIPS-Tc b) PbS-TET-CA:TIPS-Tc films. In both films, we observe interference patterns inside the QDs (dark dots) assigned to the PbS crystal packing.*

| Sample (50 mg QD:100 mg TIPS) | QD Mean (nm) | Standard deviation (nm) |
|---|---|---|
| PbS-OA | 3.9 | 0.4 |
| PbS-TET-CA | 3.9 | 0.3 |

*Table S2: PbS QDs diameter values and standard deviation as measured from the TEM images of PbS-OA:TIPS-Tc and PbS-TET-CA:TIPS-Tc films.*

# 4   Extracting an Exciton Multiplication Factor

Here, the focus is on the TIPS-Tc triplet exciton photo-physics in the TIPS-Tc, thus the following description neglects the effects of a possible intermediate state facilitating the transfer between the SF-host and QD. The overall SF-PM efficiency is given by,[13]

$$\frac{\eta_{PM}(\lambda)}{\eta_{QD}} = \alpha_{QD}(\lambda) + \alpha_{TIPS-Tc}(\lambda)\eta_{EMF} .$$

*(S1 )*

Where $\eta_{PM}$ and $\eta_{QD}$ are the IR photoluminescence quantum efficiencies (PLQE) when exciting the SF material and QD, respectively. $\alpha_i$ is the fractional absorption of component $i$. To apply this model for the triplet harvesting the IR PLQEs under 515 nm and 658 nm excitation were measured.[11] Due to the disordered and polycrystalline nature of the TIPS-Tc films, we assume the fractional absorption of each component, $\alpha_i$, are well approximated by the previously measured values in the solution phase, Figure S4.[11] We calculate the exciton multiplication factor $\eta_{EMF}$ by rearranging equation (S1 ) for $\eta_{EMF}$. For the PbS-TET-CA:TIPS-Tc film we measure values of $\eta_{QD}$ = (15.4 ± 1.0) % (excitation at 658 nm, QD only) and $\eta_{PM}(\lambda = 515\ nm)$ = (24.5 ± 1.0) % (excitation at 515 nm, QD + SF host). Taking the values for $\alpha_{QD}(\lambda)$ = 0.31 and $\alpha_{TIPS-Tc}(\lambda)$ = 0.69 at 515 nm and calculating the propagation of uncertainties, leads to $\eta_{EMF}$ = (186 ± 18) %.[11] The measurement under the TIPS-Tc triplet decay rate by ns-TA gives us a triplet harvesting efficiency of $\eta_{TET} = (97 \pm 11)$ % (section 7). Using $\eta_{EMF} = \eta_{SF}\eta_{TET}$ we find a singlet fission yield of $\eta_{SF} = (192 \pm 28)$ %.

In contrast, for the PbS-OA:TIPS-Tc film we measure values of $\eta_{QD} = (17.2 \pm 1.0)$ % and $\eta_{PM}(\lambda = 515\ nm) = (3.8 \pm 1.0)$ %. Taking the values for $\alpha_{QD}(\lambda) = 0.25$ and $\alpha_{TIPS-Tc}(\lambda) = 0.75$ at 515 nm and calculating the propagation of uncertainties, leads to $\eta_{EMF} = (-4 \pm 8)$ %.[11] Which indicates that there is essentially no triplet harvesting occurring in the PbS-OA:TIPS-Tc film.

## 5 Absorbance Spectra

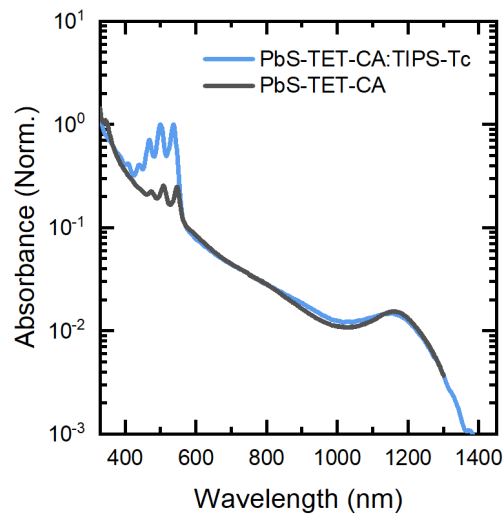

*Figure S4: Normalised absorbance spectra of a PbS-TET-CA:TIPS-Tc film (blue) and a solution of PbS-TET-CA QDs in toluene.*

## 6 Picosecond Transient Absorption

ps-Transient absorption (TA) was done on TIPS-Tc films with and without PbS-TET-CA quantum dots. The kinetics and spectral evolution of TIPS-Tc features are very similar in both cases, indicating no effect of the presence of quantum dots on the SF dynamics. The small difference is mainly due to the underlying distortion of the quantum dot dynamics which inevitably are excited at the same wavelength (535 nm).

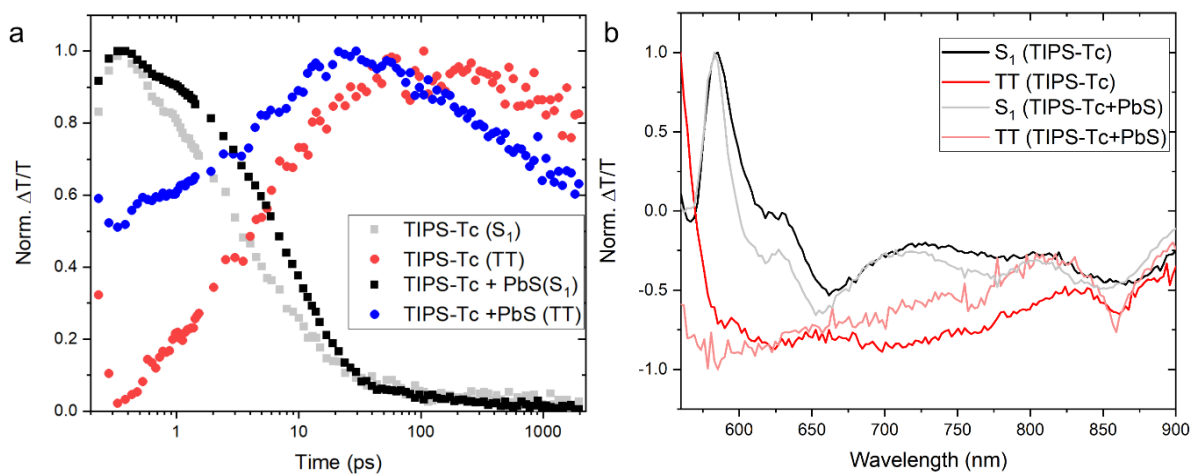

*Figure S5: a) Normalised transient absorption kinetics and b) spectral shape, comparing TIPS-Tc films with PbS-TET-CA:TIPS-Tc films. The kinetics in a) is the raw data extracted at 580 nm (S1) and 870 nm (TT) where the singlet and TT features are dominant. Therefore, any overlapping quantum dot signal will affect the kinetic trace. Still, only small changes, mainly explained by the overlapping quantum dot features, indicate that the singlet-fission process is unperturbed by the addition of quantum dots.*

# 7   Nanosecond Transient Absorption

Figures S6-7 show in the measured nanosecond TA of TIPS-Tc, PbS-TET-CA:TIPS-Tc and PbS-OA:TIPS-Tc films, we observe TIPS-Tc triplet photo induced absorption features (PIA) at ~860 and ~970 nm, after excitation of the TIPS-Tc (515 nm). These triplet PIA features are significantly quenched in the PbS-TET-CA:TIPS-Tc films, an indication of triplet transfer (Figure 4 main text). Fitting of the triplet PIA kinetics at ~970 nm with mono-exponential decays allows the extraction of the triplet intrinsic decay rate and the triplet transfer rate (Table S2). For PbS-TET-CA:TIPS-Tc films we extraction a triplet transfer rate of 0.34 ± 0.03 µs$^{-1}$, which corresponds to a triplet transfer efficiency of (97 ± 11) %.

Due to the overlapping absorption of the TIPS-Tc and PbS QD absorption excitation with 515 nm pump leads to excitation of the QD directly, as observed by the positive nsTA feature assigned to QD ground state bleach (GSB) at ~1150 nm. A further consequence of this low absorption contrast is that the comparison of the QD GSB decay under 515 and 650 nm excitation does not show a significant difference in lifetime. As an alternative, transient IR PL is used to gain insight into the transfer of triplet exciton to the PbS QDs (section 9). Figure S8 shows fitting of the QD GSB with mono-exponential decays suggesting 72 ± 2 ns and 306 ± 14 ns lifetimes of the PbS-OA and PbS-TET-CA QDs in the PbS:TIPS-Tc films (intrinsic response, 650 nm excitation).

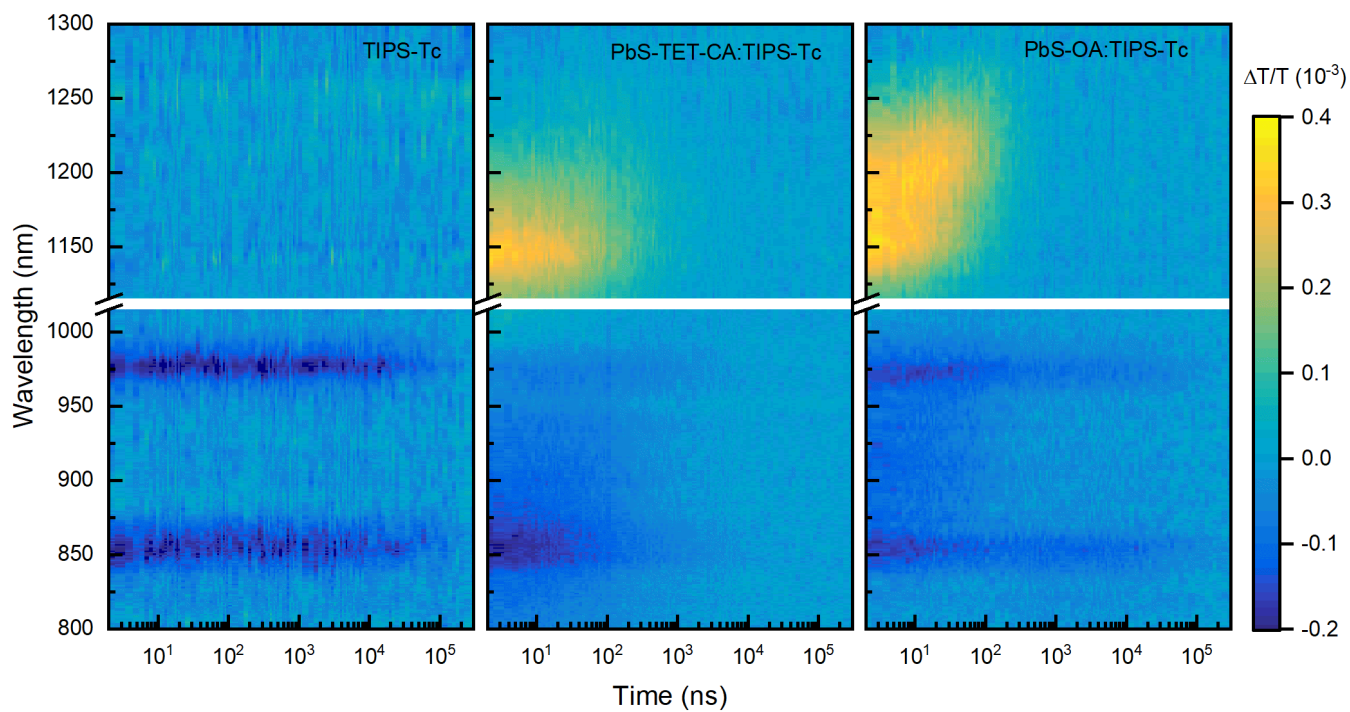

*Figure S6: Nanosecond transient absorption maps for films of TIPS-Tc, either pristine, with PbS-TET-CA or PbS-OA QDs, excited at 515 nm with ~15 µJ/cm². The TIPS-Tc triplet PIA peaks at 850 and 970 nm are clear in all cases. However, the triplet lifetime varies between the films, the PbS-Tet_CA:TIPS-Tc film having the shortest lifetime. Predominately due to direct photoexcitation, in the SF-PM systems the QD GSB is observed at 1100-1250 nm from early times (<2 ns). In the PbS-OA:TIPS-Tc there is clear red-shifting of the QD GSB within the first 10 ns after photoexcitation.*

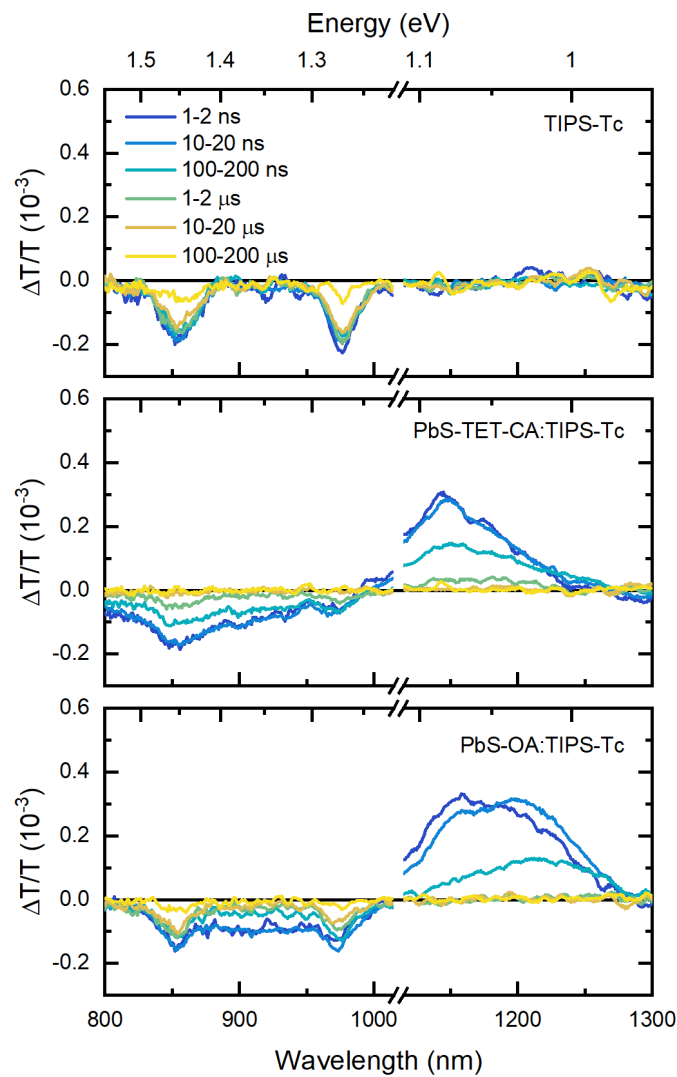

*Figure S7: Nanosecond transient absorption spectra for films of TIPS-Tc, either pristine, with PbS-TET-CA or PbS-OA QDs, excited at 515 nm with ~15 µJ/cm$^2$. Transient absorption spectra are averaged over the time ranges indicated. The TIPS-Tc triplet PIA peaks at 850 and 970 nm are clear in all films. Predominately due to direct photoexcitation, in the SF-PM systems the QD GSB is observed at 1100-1250 nm from early times (<2 ns). In the PbS-OA:TIPS-Tc there is clear red-shifting of the QD GSB within the first 10 ns after photoexcitation.*

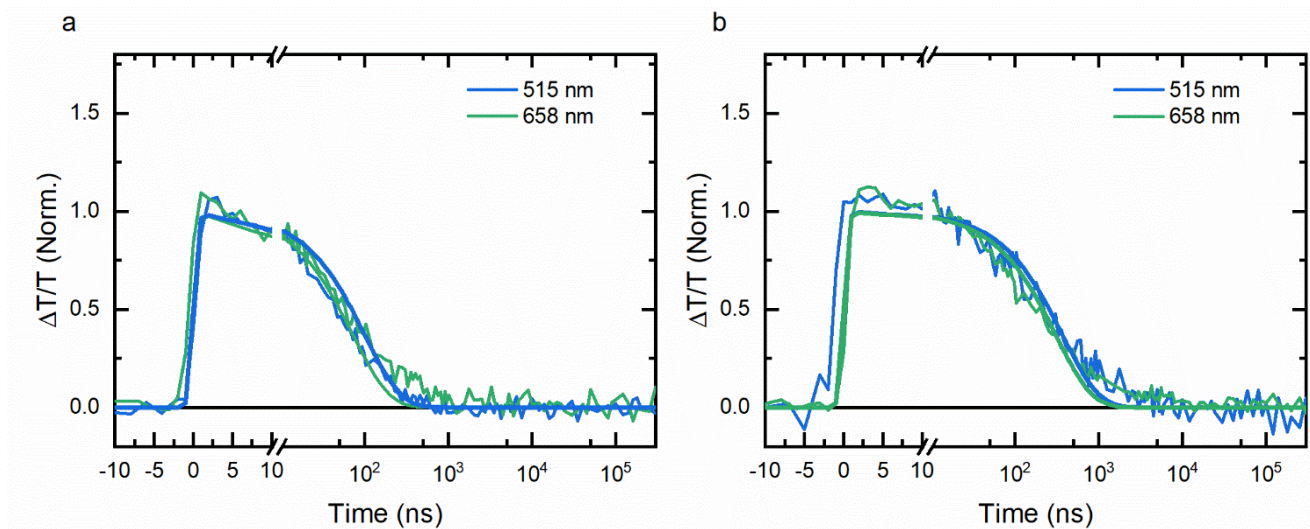

*Figure S8: Nanosecond transient absorption PbS QD GSB kinetics (1120-1180 nm) for films of PbS-OA:TIPS-Tc (a) and PbS-TET-CA:TIPS-Tc (b), excited at either 515 or 658 nm with ~15 μJ/cm². The kinetics have been fit with mono-exponential decays. The PbS-OA GSB decays with a 100 ± 6 ns and 72 ± 2 ns lifetimes when excited at 515 and 658 nm respectively. The PbS-TET-CA GSB decays with a 370 ± 25 ns and 306 ± 14 ns lifetimes when excited at 515 and 658 nm respectively.*

## 7.1 Analysis of the QD GSB Shifting

We monitor the QD GSB to track the relaxation of QD excited state population to low energy sites. As the excited states transfer to lower energy QD sites the GSB in nsTA is red-shifted. Extraction of the peak position of the GSB allows tracking of the excited state relaxation (Figure S9). We convert the wavelength shift to an energy change and fit the shift in energy to a mono-exponential decay. The PbS-OA QDs show a dramatically faster and larger drop in the energy of the QD excited states, indicating a higher degree of QD to QD transfer associated with aggregation.

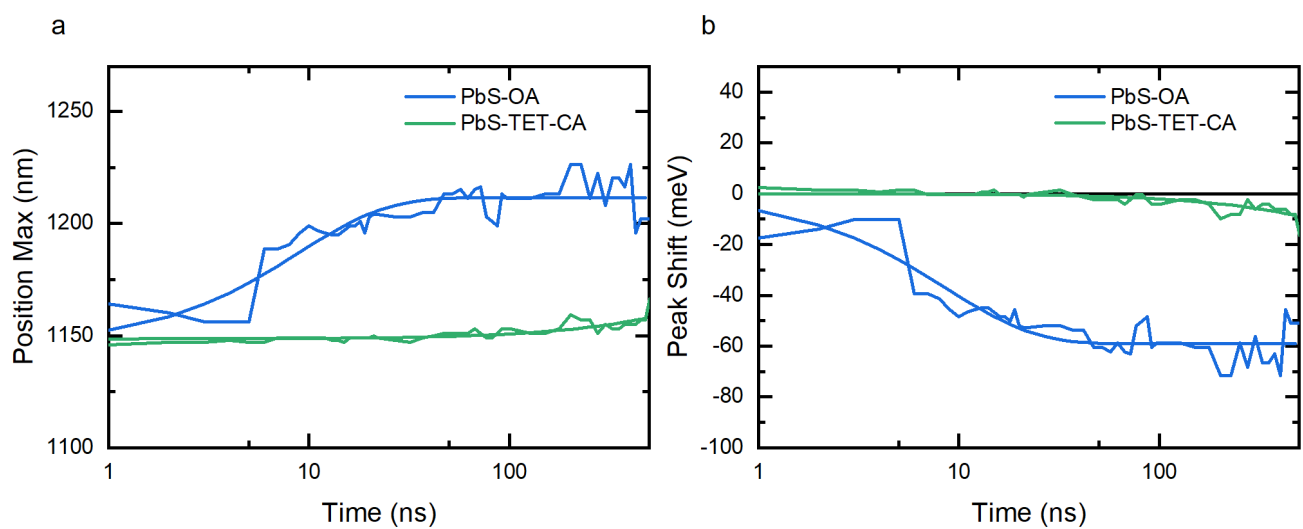

*Figure S9: Wavelength of the maximum signal for the GSB of PbS QD in films of PbS-OA:TIPS-Tc and PbS-TET-CA:TIPS-Tc after excitation with 658 nm at ~15 µJ/cm$^2$. a) peak position (b) and relative peak position shift for the PbS-OA and PbS-TET-CA GSB after 658 nm pump excitation. The shifts in peak wavelength have been parameterised with an exponential decay with offset. The peaks of the PbS-OA QD GSB drops by 60 ± 10 meV, with a time constant of 9 ± 1 ns. While the peaks of the PbS-TET-CA QD GSB drops by 10 ± 5 meV, with a time constant of 1500 ± 200 ns.*

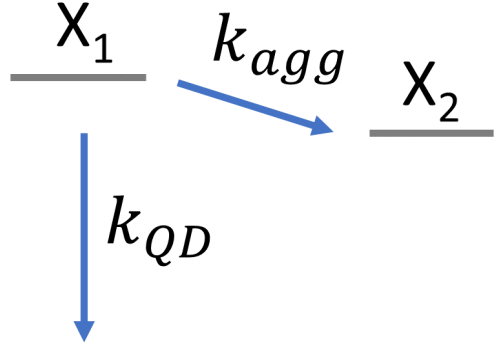

*Figure S10: Competition between exciton decay and transfer to low energy QD sites. Illustration off the simple kinetic model used to capture the branching between aggregation assisted hoping to lower QD sites and isolated QD decay rate.*

We construct a kinetic model that assumes a simplistic branching between QD excited state decay and transfer to a low energy QD site (Figure S10). Within this model, we estimate the QD aggregation assisted trapping efficiency as,

$$\eta_{agg} = \frac{k_{agg}}{k_{agg} + k_{QD}} \ . \tag{S2}$$

Taking the rates from the fitting to the position of the QD GSB peak and the decay of the total area under the GSB we calculate assisted trapping efficiencies of (90 ± 20) % and (17 ± 2)% for the OA and TET-CA capped QDs respectively.

### 7.2 Kinetic Parameters Extracted from nsTA

| Measurement | $k_1$ (1/us) | $k_{TET}$ (1/us) | $\eta_{TET}$ (%) | $k_{QD}$ (1/us) | $k_{agg}$ (1/us) | $\eta_{agg}$ (%) |
|---|---|---|---|---|---|---|
|  | TA fit | TA fit | TA fit | TA fit | TA fit | TA fit |
| TIPS-Tc | (1.05 ± 0.10)x10-2 | - | - | - | - | - |
| PbS-TET-CA:TIPS-Tc | (1.05 ± 0.10)x10-2 | 0.34 ± 0.03 | 0.97 ± 0.11 | 3.27 ± 0.15 | 0.68 ± 0.08 | 17 ± 2 |
| PbS-OA:TIPS-Tc | (1.05 ± 0.10)x10-2 | 0.0028 ±0.0016 | 0.20 ± 0.15 | 13.8 ± 0.4 | 110 ± 20 | 90 ± 20 |

*Table S3:* Kinetic parameters obtained from fitting nanosecond transient absorption kinetics. Triplet intrinsic and transfer rates are calculated from fitting of mono-exponential functions to the TIPS-Tc triplet PIA at 865-980 nm, under 515 nm excitation at ~15 µJ/cm². While the QD intrinsic decay rate is established from a mono-exponential fit to the GSB at 1120-1180 nm, under 658 nm excitation at ~15 µJ/cm². The triplet exciton transfer efficiency, $\eta_{TET}$, is calculated from the ratio between the triplet transfer rate and the sum of all relevant triplet decay channels.

## 8 Magnetic Field Dependent PL

Direct excitation of the QD in PbS-TET-CA:TIPS-Tc films with 658 nm laser light results in no observed magnetic dependence (for fields less than 0.5 T), similar to previous observations.[11, 12, 14] Under 532 nm excitation the TIPS-Tc singlet emission shows an increased PL on the application of high magnetic fields (>0.3 T), as expected for a singlet state undergoing singlet fission (Figure 3c). While, the PbS-TET-CA IR PL shows a corresponding decrease, indicating that the excited QD states are the result of triplets generated by singlet fission, transferred from the TIPS-Tc.[11, 12, 15]

## 9   IR Transient PL

We measure an instrument response function (IRF) with a full width half maximum of 5.5 ± 0.5 ns for the IR transient PL setup (figure S11). This IRF is shorter by 2 orders of magnitude than any of the time constants we observe in the triplet transfer processes. As such we treat the IRF as instantaneous relative to the dominate triplet transfer processes.

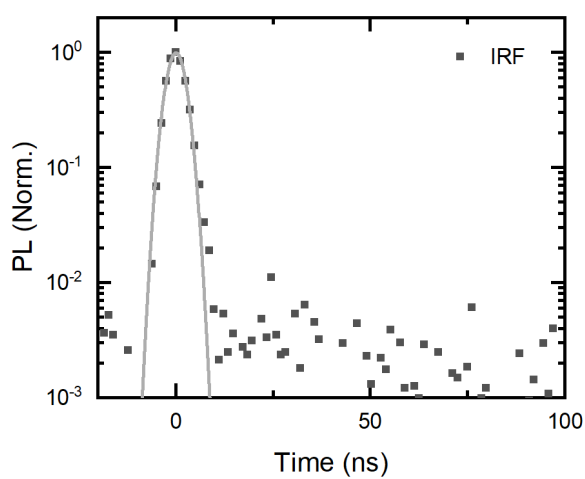

*Figure S 11: IR TCSPC instrument response function (IRF). Collected using a scattering glass substrate with 650 nm laser scatter. Fitting the IRF with a gaussian peak function is obtained with a full width half maximum of 5.5 ± 0.5 ns.*

With the laser excitation blocked before entering the sample area, the detector counts due to ambient conditions were measured for the same exposure time as the transient PL measurements. The mean detector counts per time bin, across the time window, represents the background PL counts. The PL kinetics were correct by subtraction of this value. After subtraction of this background value, positive values for the PL kinetics at times before the laser pulse are observed. These PL levels (t < 0) arise due to the periodic nature of the experiment. Periodicity in the PL kinetics is explicitly included in the time-series deconvolution analysis (see below) due to the periodicity of the fast Fourier transform. We perform fluence-dependent measurements to investigate the effect of any non-linear behaviour of the QD excited state (650 nm excitation) or TIPS-Tc triplet decay (535 nm excitation) (Figure S12). Over the range of incident fluences investigated, we observed no dependence of the transient IR PL decay, indicating even at the highest fluence used 15 nJ/cm$^2$ the system is in the low excitation density regime where bimolecular decay in the QD or the TIPS-Tc triplet can be ignored. While the IR PL shows an extended lifetime under excitation of the TIPS-Tc (535 nm) relative to excitation of the QD alone (650 nm) in the PbS-TET-CA:TIPS-Tc film (Figure 4 main text), the PbS-OA:TIPS-Tc does not show any significant extension (Figure S13). In agreement with the steady-state observation that effectively no triplet transfer is occurring. We extract QD excited state decay rates of 2.3 ± 0.2 µs$^{-1}$ and 2.5 ± 0.2 µs$^{-1}$ by fitting the transient IR PL decay of PbS-OA:TIPS-Tc and PbS-TET-CA:TIPS-Tc respectively, under 650 nm excitation (Figure S14).

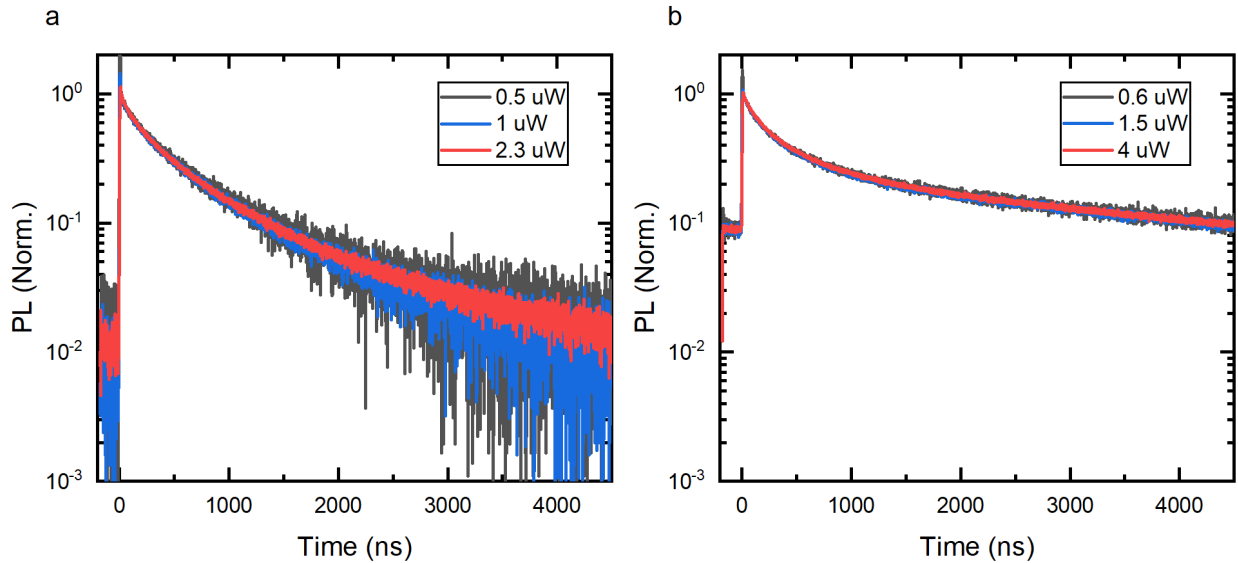

*Figure S12: Normalised IR transient PL kinetics for a PbS-TET-CA:TIPS-Tc film. The PbS-TET-CA:TIPS-Tc film was excited at 650 nm (a) and 535 nm (b) with varying fluences. 650 nm excitation at ~2.5, 5, 10 nJ/cm², was used. 535 nm excitation at ~2, 7 and 15 nJ/cm², at 0.2 MHz repetition rate was used. Contribution to the detected counts by background counts was removed before normalisation to the initial value of the PL decay.*

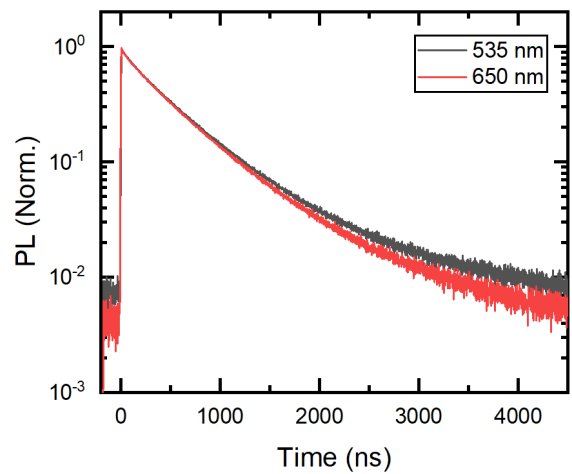

*Figure S13: Normalised IR Transient PL Kinetics. The PbS-OA:TIPS-Tc film was excited at 650 nm (2.3 µW, 10 nJ/cm²) and 535 nm (4.0 µW, 15 nJ/cm²), at 0.2 MHz repetition rate. Contribution to the detected counts by background counts was removed before normalisation to the initial value of the PL decay.*

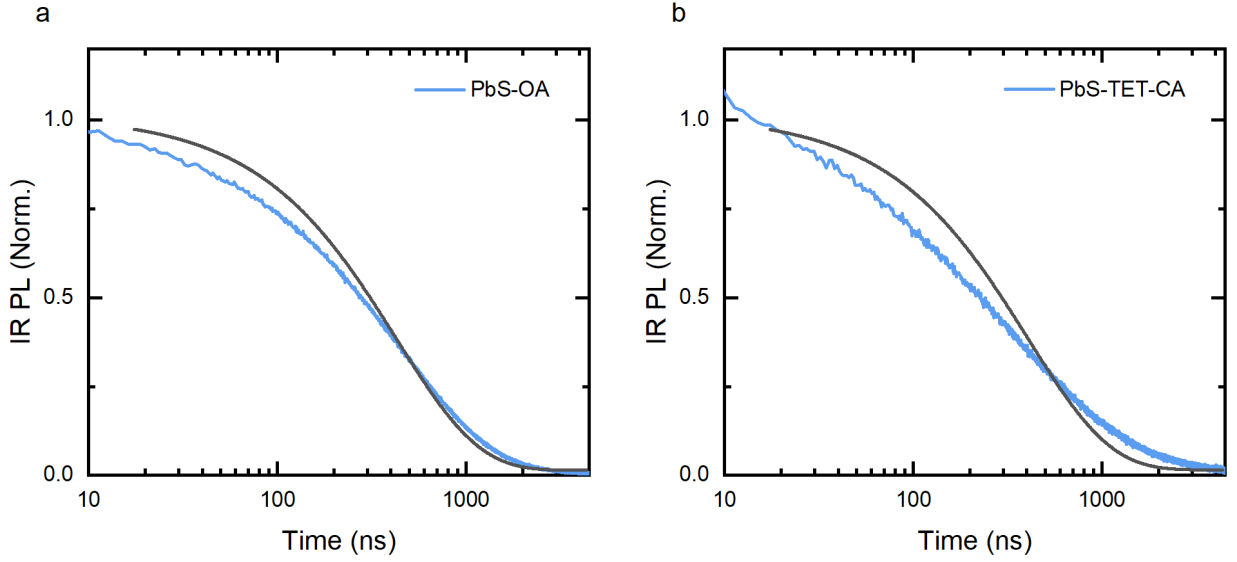

*Figure S14: Normalised IR transient PL kinetics for films of a) PbS-OA:TIPS-Tc and b) PbS-TET-CA:TIPS-Tc excited at 650 nm with fluence 10 nJ/cm², 0.2 MHz repetition rate (QD preferentially excited). The PL decay was fitted with a single exponential decay with decay rates 2.3 ± 0.2 µs⁻¹ and 2.5 ± 0.2 µs⁻¹ respectively.*

## 9.1 Triplet Flux Deconvolution

The flux (assigned to triplet transfer) into the PbS QD, $\phi_T(t)$, is found by deconvolving the intrinsic QD response $h(t)$ (650 nm excitation, the QDs impulse response) from the observed QD response with triplet transfer $y(t)$ (535 nm excitation). Here the ansatz is that the QD dynamics can be related to the intrinsic response as follows,[15]

$$y(t) = h(t) * \left(\delta(t) + \phi_T(t)\right). \tag{S3}$$

Where $\delta(t)$ is a delta function representing the fraction of photons in the 535 nm pump pulse that excites the QDs directly. To achieve appropriate levels of signal to noise, we perform a post-processing step where the $y(t)$ and $h(t)$ time series are binned, taking the average of 40 data points and reducing it to 1 data point respectively (at the mean time of the 40 data points). The deconvolution is calculated using a fast Fourier transform (FFT) as described by,

$$\delta(t) + \phi_T(t) = FFT^{-1}\left[\frac{FFT[y(t)](\omega)}{FFT[h(t)](\omega)}\right](t). \tag{S4}$$

The first 2 two data points after the pump excitation were removed (removing $\delta(t)$) giving the triplet flux into the PbS QDs (Figure S15b). This triplet flux is then convolved with the intrinsic QD decay $h(t)$ to give the QD IR PL (directly proportional to the QD population) that is due to triplet transfer (Figure S15c). The triplet flux shows unexpected behaviour, where it rises over the first ~500 ns after the pump pulse. In previous measurements on bilayers of tetracene and PbS QDs, the deconvolution of the QD PL showed a triplet flux that doesn't rise at all after the pump excitation and only decays over a μs time scale.[15]

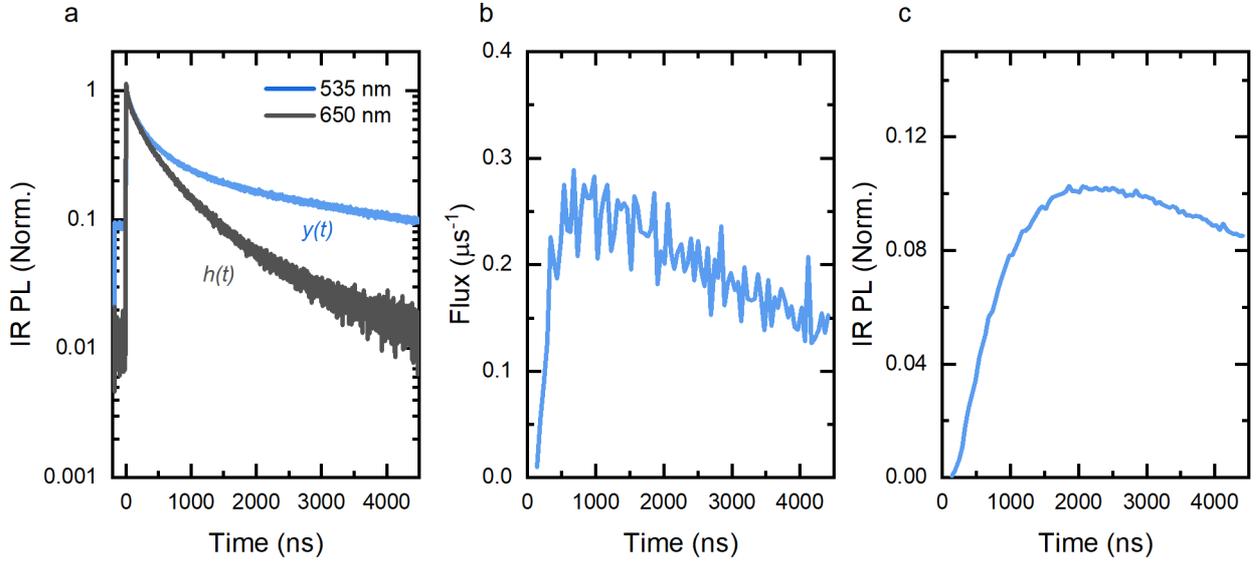

*Figure S14: a) PbS-TET-CA:TIPS-Tc transient IR PL for the PMF under intrinsic decay (650 nm excitation) and triplet transfer (535 nm excitation), at 0.2 MHz repetition rate. b) Deconvoluted excitation flux into the PbS QD). c) Triplet flux convolved with the intrinsic QD decay (h(t)) to give the PbS PL resulting from purely triplet transfer.*

## 9.2  Triplet Transfer Models

Here we discuss possible kinetics schemes for the triplet transfer between TIPS-Tc and PbS QDs and compare the dynamics to the IR transient PL measurements. We propose two kinetic models, the first requires only two species, the TIPS-Tc triplet $[T_{1,1}]$ and the excited QD state $[QD_1]$; the second introduces a third intermediate species $[I]$.

### 9.2.1  Two-Species Model

In this kinetic scheme we assume there are only two states participating in the triplet transfer, the TIPS-Tc triplet state $[T_{1,1}]$ and the excited PbS QD state $[QD_1]$. The dynamics of this system are described as follows,

$$\frac{d[T_{1,1}]}{dt} = -(k_1 + k_{TET})[T_{1,1}] - k_2[T_{1,1}]^2 + \eta_{SF} \cdot G_T(z),$$

(S5)

$$\frac{d[QD_1]}{dt} = -k_{QD}[QD_1] + k_{TET}[T_{1,1}],$$

(S6)

with rates as described earlier. To simplify this system of differential equations we assume the case of the low $[T]$ limit where contribution by the $k_2[T]^2$ can be ignored and that triplet transfer out-competes triplet intrinsic decay ($k_1 + k_{TET} \sim k_{TET}$). Solving this system leads to a triplet population given by,

$$[T_{1,1}](t) = [T_{1,1}]_0 e^{-k_{TET}t}, \qquad (S7)$$

where $[T_{1,1}]_0$ is the initial triplet density after singlet fission. The triplet flux into the QD is,

$$\phi_T(t) = k_{TET}[T_{1,1}](t) = k_{TET}[T_{1,1}]_0 e^{-k_{TET}t}. \qquad (S8)$$

The QD population due to this transfer is given by,

$$[QD_1]_\phi(t) = -\frac{k_{TET}[T_{1,1}]_0}{k_{TET} - k_{QD}}(e^{-k_{TET}t} - e^{-k_{QD}t}). \qquad (S9)$$

While the QD population due to direct excitation (650 nm excitation) is,

$$[QD_1]_{PL}(t) = [QD_1]_0 e^{-k_{QD}t}, \qquad (S10)$$

where $[QD_1]_0$ is the initial excited QD population. This set of equations allows for simultaneous calculation of the QD intrinsic PL decay $[QD_1]_{PL}(t)$, the triplet flux into the QD $\phi_T(t)$ and QD population due to transfer $[QD_1]_\phi(t)$. Figure S16 shows the best achieved global fitting of these functions to the measured values. The quality of this fit is very poor, showing large systematic discrepancies of the observed trends. Notably, the triplet flux does rise over the first ~500 ns as observed in the measured data and the QD population from transfer peaks and falls faster than measured. We constrain the system such that the triplet transfer rate is the same as given by the ns-TA measurements. The value for the QD intrinsic decay rate $k_{QD}$ is slightly smaller than the values measured by the fitting of the QD GSB in ns-TA and the observed transient PL decay (Table S4).

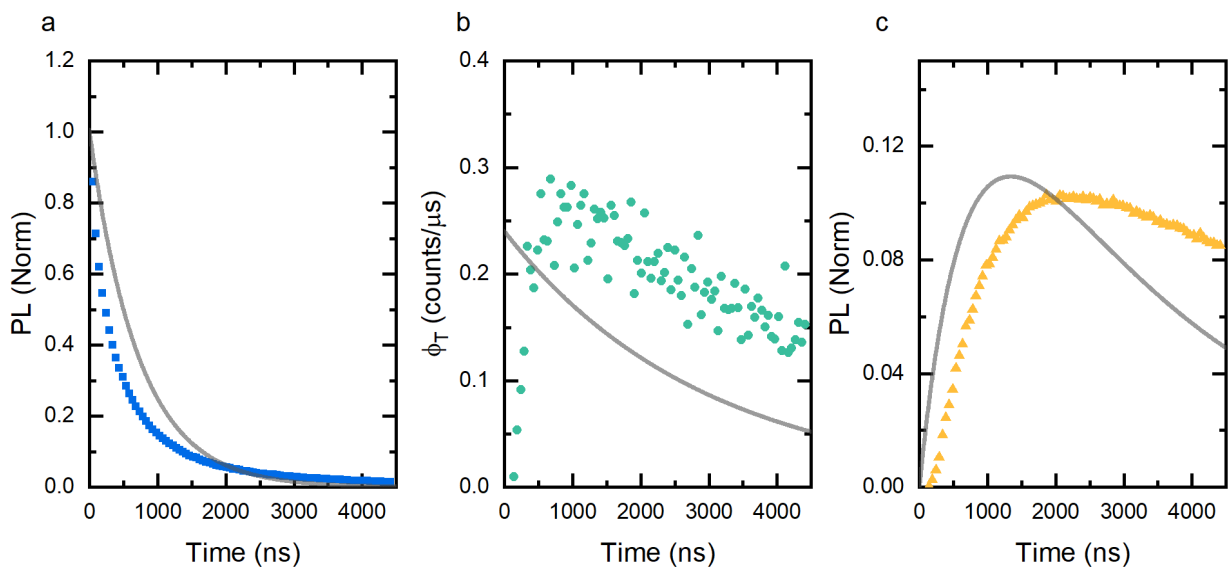

*Figure S15: Two species model fitting of the IR transient PL for a PbS-TET-CA:TIPS-Tc film. The intrinsic QD PL decay, triplet flux into the QD and QD PL counts from triplet transfer, was fitted globally. a) PbS-TET-CA:TIPS-Tc intrinsic QD PL decay (650 nm excitation). b) Triplet flux into the PbS QD in a film of PbS-TET-CA:TIPS-Tc after 535 nm excitation of the SF-host. c) PbS-TET-CA:TIPS-Tc QD PL resulting from triplet transfer (after excitation with 535 nm), calculated by convolution of the triplet flux and the PbS-TET-CA intrinsic decay dynamics.*

| $k_{TET}$ (1/µs) | $k_{QD}$ (1/µs) | $k_{QD}$ (1/µs) | $k_{QD}$ (1/µs) |
|---|---|---|---|
| TA fit | TA fit | TrPL fit | Transfer TrPL fit |
| 0.34 ± 0.03 | 3.3 ± 0.2 | 2.5 ± 0.2 | 1.4 ± 0.1 |

*Table S4: Comparison of the triplet transfer kinetic parameters for a two-species model with global fitting to the intrinsic QD decay, triplet flux and QD PL from triplet transfer.*

Figure S17 shows the best achieved two species fitting against the measured QD PL from transfer alone (not globally fitted). The fitted kinetic for the transient PL from triplet transfer is reasonable, showing a lower discrepancy with the measured response. However, the corresponding kinetics for the QD intrinsic decay and triplet flux show considerable deviation from the data. This fitting method requires a significantly slower QD decay relative to the values measured by nsTA and transient PL alone (Table S5). This discrepancy between QD decay rate extracted by the QD population from triplet transfer (535 nm excitation) and the rate obtained by optical excitation of the QDs directly (650 nm excitation) is consistent with the hypothesis that there exist two subsets of QDs within the film. One set that is affected by QD aggregation to a greater extent, resulting in short QD lifetimes due to trapping and lower triplet transfer due to the separation of triplet donor and acceptor. The other subset of QDs are isolated within the SF-host having slower decay (similar to the rate measured for an isolated dot in solution ~0.5 µs$^{-1}$) and high triplet transfer due to the maximal interaction between donor and acceptor. The rise in triplet flux could be an artefact in this case as the ansatz in equation S3 would not be valid. We leave the investigation of this hypothesis to future work.

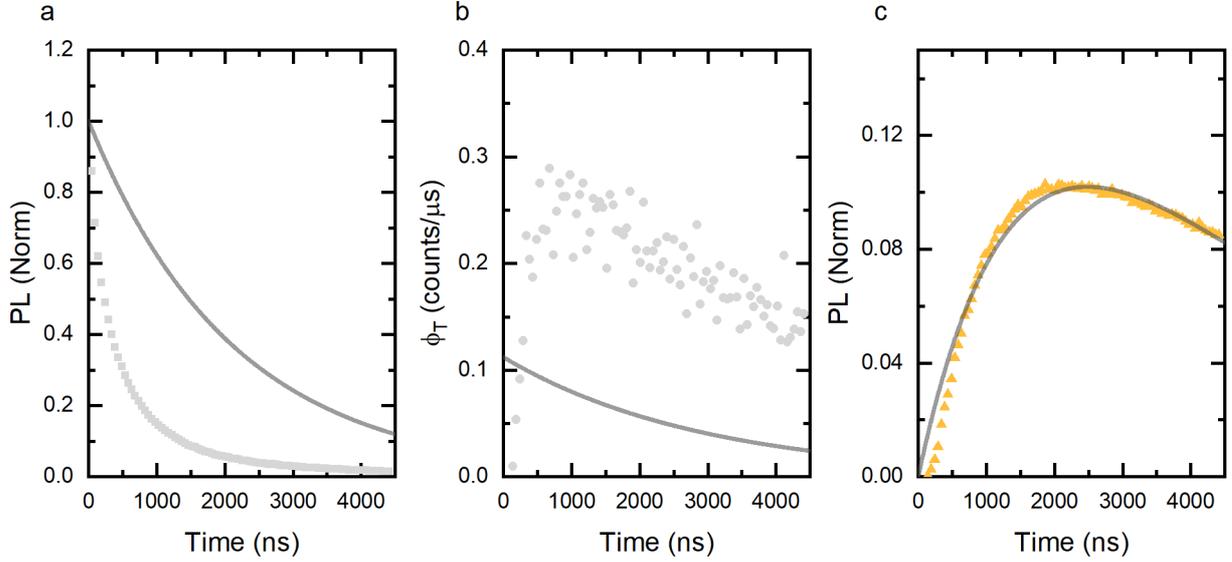

*Figure S16: Two species model fitting of the IR transient PL for a PbS-TET-CA:TIPS-Tc film. Only data for QD PL from transfer was fitted and the required intrinsic QD PL decay, triplet flux into the QD plotted beside the measured counts. a) PbS-TET-CA:TIPS-Tc intrinsic QD PL decay (650 nm excitation). b) Triplet flux into the PbS QD in a film of PbS-TET-CA:TIPS-Tc after 535 nm excitation of the SF-host. c) PbS-TET-CA:TIPS-Tc QD PL resulting from triplet transfer (after excitation with 535 nm), calculated by convolution of the triplet flux and the PbS-TET-CA intrinsic decay dynamics.*

| $k_{TET}$ (1/μs) | $k_{QD}$ (1/μs) | $k_{QD}$ (1/μs) | $k_{QD}$ (1/μs) |
|---|---|---|---|
| TA fit | TA fit | TrPL fit | Transfer TrPL fit |
| 0.34 ± 0.03 | 3.27 ± 0.15 | 2.5 ± 0.2 | 0.50 ± 0.05 |

*Table S5: Comparison of the triplet transfer kinetic parameters for a two-species model with the fitting of the QD PL from transfer alone.*

### 9.2.2 Three-Species Model

In this kinetic scheme, we assume the existence of an intermediate state $[I]$ participating in the triplet transfer. The dynamics of this system are described as follows,

$$\frac{d[T_{1,1}]}{dt} = -(k_1 + k_{TET_1})[T_{1,1}] - k_2[T_{1,1}]^2 + \eta_{SF} \cdot G_T(z),$$

( S11 )

$$\frac{d[I]}{dt} = -k_{TET_2}[I] + k_{TET_1}[T_{1,1}],$$

( S12 )

$$\frac{d[QD_1]}{dt} = -k_{QD}[QD_1] + k_{TET_2}[I],$$

( S13 )

where $k_{TET_1}$ is the triplet transfer rate from the TIPS-Tc to the intermediate (this rate is the same as the previously discussed $k_{TET}$ as it quantifies triplet loss from the TIPS-Tc) and $k_{TET_2}$ is the rate of triplet transfer from the intermediate to the QD excited state. To simplify this system of differential equations we assume the case of the low $[T]$ limit where contribution by the $k_2[T]^2$ can be ignored and that triplet transfer out-competes triplet intrinsic decay ($k_1 + k_{TET_1} \sim k_{TET_1}$) and there is 100% transfer from the intermediate to the QD). Solving this system leads to a triplet population given by,

$$[T_{1,1}](t) = [T_{1,1}]_0 e^{-k_{TET_1} t},$$

(S14)

where $[T_{1,1}]_0$ is the initial triplet density after singlet fission. The intermediate state population is given by,

$$[I](t) = -\frac{k_{TET_1}[T_{1,1}]_0}{k_{TET_1} - k_{TET_2}} \left( e^{-k_{TET_1} t} - e^{-k_{TET_2} t} \right),$$

(S15)

where we assume the initial population of the intermediate state is $[I](0) = 0$. The triplet flux into the QD is,

$$\phi_T(t) = k_{TET_2} I(t) = -\frac{k_{TET_1} k_{TET_2}[T_{1,1}]_0}{k_{TET_1} - k_{TET_2}} \left( e^{-k_{TET_1} t} - e^{-k_{TET_2} t} \right),$$

(S16)

The QD population due to this transfer is given by,

$$[QD_1]_\phi(t) = \frac{k_{TET_1} k_{TET_2}[T_{1,1}]_0 \left( k_{QD}\left(e^{-k_{TET_2} t} - e^{-k_{TET_1} t}\right) + k_{TET_2}\left(e^{-k_{TET_1} t} - e^{-k_{QD} t}\right) + k_{TET_1}\left(e^{-k_{QD} t} - e^{-k_{TET_2} t}\right) \right)}{(k_{TET_1} - k_{TET_2})(k_{TET_1} - k_{QD})(k_{TET_2} - k_{QD})},$$

(S17)

While the QD population due to direct excitation (650 nm excitation) is,

$$[QD_1]_{PL}(t) = [QD_1]_0 e^{-k_{QD} t},$$

(S18)

where $[QD_1]_0$ is the initial excited QD population. This set of equations allows for simultaneous calculation of the QD intrinsic PL decay $[QD_1]_{PL}(t)$, the triplet flux into the QD $\phi_T(t)$ and QD population due to transfer $[QD_1]_\phi(t)$. Figure S18 shows the best achieved global fitting of these functions to the measured values. We constrain the system such that the triplet transfer rate $k_{TET_1}$ is the same as given by the ns-TA measurements. The agreement between measured values and fit is the strongest out of the three investigated fitting procedures, reproducing the observed rises and falls in the various time-dependent quantities. To accurately fit the rise in the triplet flux requires the $k_{TET2}$ fitting parameter. This introduction of an intermediate state is not arbitrary as it has been shown that the TET-CA ligand is crucial to the triplet transfer process in solution and its rate of transfer into the PbS QD has been calculated.[16] Thus we assign this intermediate state as the TET-CA triplet $[I] = [T_{1,2}]$. For illustrative purposes, the TET-CA triplet and TIPS-Tc triplet populations in Figure 4.d (main text), have been smoothed with using the Savitzky-Golay method with a 2nd order polynomial fitting over a 10 and 7 data point window respectively.

Table S6 compares the various kinetic parameters. The value for the QD intrinsic decay rate $k_{QD}$ is again slightly smaller than the values measured by the fitting of the QD GSB in ns-TA and the observed transient PL decay. This might be to the two subset hypothesis mentioned previously.

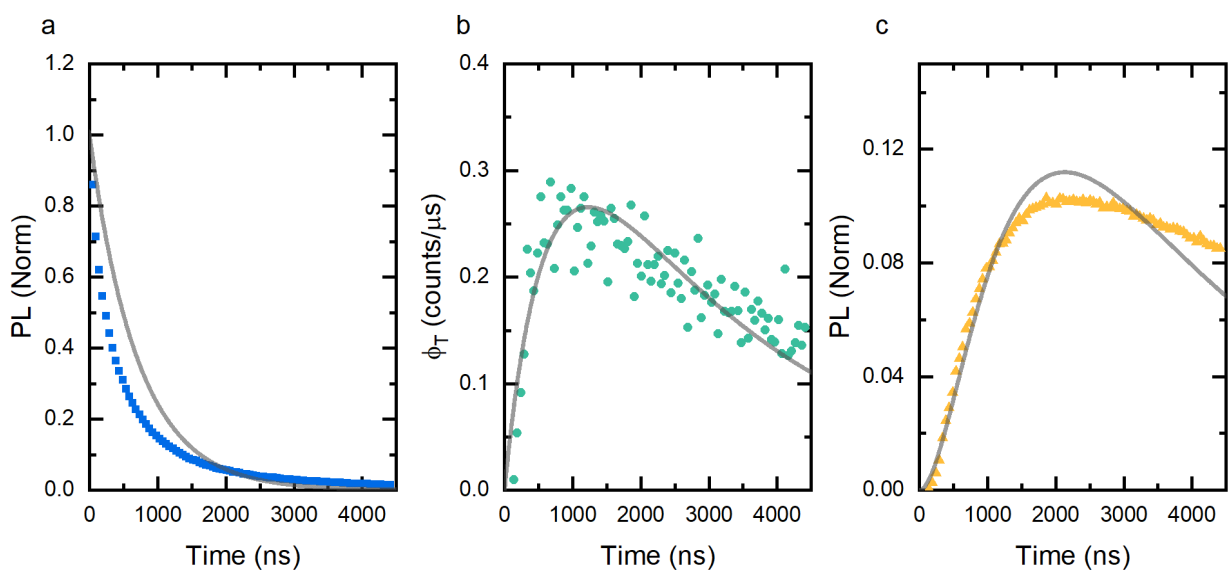

*Figure S17: Three species model fitting of the IR transient PL for a PbS-TET-CA:TIPS-Tc film. The intrinsic QD PL decay, triplet flux into the QD and QD PL counts from triplet transfer, was fitted globally. a) PbS-TET-CA:TIPS-Tc intrinsic QD PL decay (650 nm excitation). b) Triplet flux into the PbS QD in a film of PbS-TET-CA:TIPS-Tc after 535 nm excitation of the SF-host. c) PbS-TET-CA:TIPS-Tc QD PL resulting from triplet transfer (after excitation with 535 nm), calculated by convolution of the triplet flux and the PbS-TET-CA intrinsic decay dynamics.*

| $k_{TET1}$ (1/µs) | $k_{TET2}$ (1/µs) | $k_{QD}$ (1/µs) | $k_{QD}$ (1/µs) | $k_{QD}$ (1/µs) |
|---|---|---|---|---|
| TA fit | $\phi_T(t)$ fit | TA fit | TrPL fit | Transfer TrPL fit |
| 0.34 ± 0.03 | 1.6 ± 0.1 | 3.27 ± 0.15 | 2.5 ± 0.2 | 1.40 ± 0.2 |

*Table S6: Comparison of the triplet transfer kinetic parameters for a three-species model with global fitting to the intrinsic QD decay, triplet flux and QD PL from triplet transfer.*